\documentclass{article}

\usepackage{PRIMEarxiv}
\usepackage{orcidlink}
\usepackage[utf8]{inputenc} 
\usepackage[T1]{fontenc}    
\usepackage{hyperref}       
\usepackage{float}
\usepackage{url}            
\usepackage{booktabs}       
\usepackage{amsfonts}       
\usepackage{nicefrac}       
\usepackage{microtype}      
\usepackage{lipsum}
\usepackage{fancyhdr}      
\usepackage{graphicx}       
\graphicspath{{media/}}    
\usepackage{array}
\usepackage{xcolor}
\usepackage{enumitem}
\usepackage{placeins}
\usepackage{multirow}

\title{Hungary and AI\\ efforts and opportunities\\ in comparison to Singapore
}

\author{
  András Ferenczy\,\orcidlink{0009-0007-9520-1989}\\
  Student, Leadership Academy \\
  Mathias Corvinus Collegium \\
  Budapest, Hungary\\
  \texttt{aferenczy.education@gmail.com} \\
}

\begin{document}
\maketitle

\begin{abstract}
The study assesses Hungary's National AI Strategy and its implementation through the analysis of strategic documents, publicly available financial records and expert interviews with the Hungarian AI Coalition President and Chief Strategic Advisor to the Government Commissioner for AI. 
22 goals from Hungary's strategy were evaluated through conceptual, governance, temporal, and financial dimensions before being benchmarked against Singapore’s National AI Strategies (NAIS 1.0 and NAIS 2.0). Key findings include an estimated total of EUR 4.65 billion in AI-related public investment in Hungary. Openly available financial data was found for only half of the evaluated goals, and just three projects made up 98\% of all documented funding. The research also reveals Hungary’s implementation challenges, including fragmented execution following ministerial reorganizations and the absence of designated biennial reviews since 2020.
Furthermore, the paper provides targeted recommendations for Hungary’s forthcoming AI strategy, drawing on Singapore’s framework as a reference point. These include adapting to the era of large language models, restructuring the existing triple helix network to foster more effective dialogue and advocacy, and positioning the country as an East-West bridge for automotive AI experimentation.
\end{abstract}

\newpage
\section{Introduction}
\label{sec:introduction}

\subsection{Introduction to the topic}
\label{subsec:introduction-to-the-topic}

Artificial Intelligence (AI) has emerged as a pivotal driver of innovation and economic transformation globally. In response, nations have developed strategic frameworks to foster AI adoption and integration. This research examines the governmental initiatives of Hungary in comparison to the benchmark set by Singapore, a recognized global leader in AI. The study explores how effectively Hungary leverages AI-driven growth relative to this international reference point and identifies key areas and factors where further improvement may be possible.

\subsection{Research background}
\label{subsubsec:research-background}

Starting from the 1990s, Hungary’s economic development has become increasingly dependent on innovation and technological advancement because services already account for 57.19\% of national exports\cite{oneill2024}.  
Despite recent progress, Hungary still occupies only a middle position in global AI rankings\cite{tortoise2024} and is outperformed by several emerging economies.  
Singapore, in contrast, ranks among the top countries world-wide for AI adoption, innovation, and infrastructure\cite{aiindex2024}, making it an ideal benchmark for comparison.

\subsection{Research problem}
\label{subsubsec:research-problem}
Hungary faces a substantial implementation gap in translating AI policy into measurable economic benefits.  
The core research question therefore asks:  
How effectively does Hungary’s National AI Strategy\cite{euhungary2023} support the strategic adoption and utilisation of AI to enhance economic competitiveness, and what improvements can be identified by benchmarking against Singapore's strategy\cite{nais2019}\cite{nais2023}?

\subsection{Research questions}
\label{subsec:research-questions}

\begin{itemize}
\item What strategic approaches have Hungary and Singapore adopted in their national AI strategies?
\item How effectively has Hungary implemented AI strategies based on predefined objectives and measurable indicators?
\item What are the key financial commitments in both Hungary's and Singapore's AI strategic plans?
\item What practical insights and recommendations can Hungary derive from Singapore's national AI strategies?
\end{itemize}

\subsection{Limitations}
\label{subsec:limitations}

\begin{itemize}
\item AI strategy policies are mostly new (Hungary: 2020, Singapore: 2023) and are in early stages of implementation, making it hard to measure direct impacts.
\item Regional and international comparisons may have inconsistencies due to differences in context (population, size of the country, GDP per capita, etc.)
\item The research relies primarily on publicly available online documents, press releases, and institutional publications. Internal governmental reports, unpublished updates, or confidential implementation details were not accessible, which constrains the completeness and depth of assessment, particularly in evaluating ongoing initiatives or measuring behind-the-scenes coordination. Expert interviews may unveil this information partially.
\end{itemize}

\newpage
\section{Literature review}
\label{sec:literature-review}

\subsection{Hungary and AI}
\label{subsec:hungary-ai}

An overview of the history of AI and Hungary is summarized by Sántáné-Tóth\cite{santane2007}, who examines the period from 1975 to 1996, highlighting significant developments. The beginning of AI in Hungary can be linked to László Kalmár, who designed a machine in the 1950s which could be programmed. This milestone laid ground for the start of research and education of machine learning in the country. A next achievement occurred in 1975, with the development of a Hungarian Prolog interpreter, which made Hungary worldwide famous. Eötvös Loránd University and the Technical University of Budapest consequently started their first AI related courses connected to their informatics education in the 1980s. After a longer cooldown period, 1996 was another important year for the Hungarian AI community: the European Coordinating Committee for Artificial Intelligence organized the European Conference on AI (the 12th ECAI) in Budapest.

Two decades later, with AI becoming the epicentrum of technological innovation in the 2010s, the European Union adopted a ``Coordinated Plan on Artificial Intelligence''\cite{eucoordinated2018}. In accordance with the plan, Hungary drafted its first AI strategy\cite{euhungary2020} in 2020. The strategy consists of three main focus areas: 1) strengthening the foundation pillars of the Hungarian AI ecosystem: (e.g.: research development, education and competence development, infrastructure deployment); 2) promoting the technology related fields with the highest reward potential for the country: manufacturing, healthcare, agriculture, public administration, transportation, logistics and energy; 3) initiating transformative programs that offer direct benefits to citizens (e.g.: autonomous systems and self-driving vehicles, health-consciousness programs, climate-driven agriculture solutions). In 2024, following the EU AI Act\cite{euaiact2024}, the commission launched a ``AI innovation package''\cite{euinnovation2024} to support such startups, small and medium-sized enterprises. Although the package is still under development, it is considered a huge future initiative to foster AI related startups.

The challenge of measuring the economic impact of such policies is conducted by Nxumalo et al.\cite{nxumalo2022}, who measures using a multiple linear regression model with the following indicators: Gross Domestic Product (GDP), Total Factor Productivity, Foreign Direct Investment, Net Investment of Financial Assets. The data covers a sample period of 30 years from 1991 to 2020 and concludes that an increase in the ``AI index'' by one unit increases GDP by an estimated 3.23 percent.

Focusing not only on economic impact but on many other aspects of the AI's transformative effect in Hungary such as on warfare, health, education is done by Kovács et al.\cite{kovacs2023}. The collection of papers, authored by researchers from six Hungarian universities concludes, among others, that a technology-based paradigm shift has to be done in order to properly deal with AI related challenges.

\subsection{Singapore and AI}
\label{subsec:singapore-ai}

A summary of how Singapore reached the forefront of AI research, innovation and adaptation was written by Khanal et al.\cite{khanal2024}. The paper mentions that the first step in the direction of AI was made in the 1980s, when the government launched nationwide computerisation plans (e.g. the National Computerisation Plan in 1980 and Civil Service Computerisation Programme in 1981) to digitize the economy and public sector. These early initiatives – followed by IT masterplans like IT2000 in 1992 envisioning an ``Intelligent Island'' – laid a digital foundation for later AI efforts. By the mid-2010s, Singapore pivoted explicitly towards data and AI: in 2014 the Smart Nation program was launched to harness emerging technologies and ``make the economy more productive, the lives better, and the society more responsive to people's needs'' By the mid-2010s, Singaporean government foresight groups such as the Committee on the Future Economy which had called for building strong digital capabilities – including AI – to drive growth. This set the stage for national-level AI strategies. Leveraging Singapore's strengths (robust ICT infrastructure, unified government and technology-oriented talent), policymakers envisioned using AI to transform the economy and society.

The research done by Goode et al.\cite{goode2023} also analyses Singapore's AI progress, and highlights that in May 2017, Singapore's National Research Foundation (NRF) launched AI Singapore (AISG) as a flagship national AI R\&D program. AISG was a government-wide initiative to ``catalyse, synergise and boost'' the nation's AI capabilities. The program brought together key agencies – NRF, the Smart Nation Digital Government Office (SNDGO), Economic Development Board (EDB), Infocomm Media Development Authority (IMDA), SGInnovate, healthcare IT units – as well as all Singapore-based research institutions and AI startups into a unified effort\cite{nrf2017}. An emphasis on talent development accompanied the surge in R\&D: IMDA had earlier introduced an AI Apprenticeship Programme (AIAP) to reskill professionals in machine learning; by 2018, the third batch of AIAP trainees had graduated. Universities responded by launching AI-focused courses and even specialized centers. Notably, Nanyang Technological University was ranked the world's top university for AI research citation impact in 2017, underscoring the strength of Singapore's academic contributions. These efforts to train and attract skilled AI researchers helped address the talent gap – a critical challenge given an anticipated demand of 60,000 ICT professionals by 2024.

At this time, concurrently, Singapore also took proactive steps to address AI ethics and governance at a national level, as discussed by Chia\cite{chia2018}. In 2018, the government convened the Advisory Council on the Ethical Use of AI and Data, a multi-stakeholder panel including industry leaders (from Google, Microsoft, Alibaba, local firms). The council's mandate was to advise on responsible AI development and use, helping to draft guidelines and frameworks for ethical AI deployment. At the same time, IMDA launched a five-year research program on AI governance, led by Singapore Management University, to study legal, policy, and societal implications of AI. These efforts resulted in Singapore releasing the Model AI Governance Framework in January 2019 – one of the world's first national frameworks providing detailed, implementable guidance for private sector AI ethics.

The NAIS 1.0 (published also in 2019) has also been evaluated by Khanal et al. and observed that the Singaporean model stands out for ``combining centralized strategic control with decentralized sectoral experimentation.'' As a conclusion, the paper states that with NAIS 2.0, efforts have enabled Singapore to emerge as one of the global leaders in AI development.

\subsection{Singapore and Hungary}
\label{subsec:singapore-hungary}

There is only a few research dedicated to comparing Singapore and Hungary in any aspect. The publicly available research includes György and Veress's\cite{gyorgyveress2013} paper, which compare the external financial vulnerabilities of Hungary and Singapore as small, open economies. They conclude that Singapore achieved the gains of economic openness with much lower external debt risks than Hungary, thanks to prudent policies in areas like foreign-account liberalization and public financing. In contrast, Hungary's post-2008 difficulties with external debt highlighted its financial fragility, suggesting that adopting elements of Singapore's policy approach could help reduce Hungary's vulnerability.

Another research written by György and Sebestyén\cite{gyorgysebestyen2013} provide a comparative study of minority education initiatives, examining Singapore's success in closing the ``Malay problem'' achievement gap and drawing lessons for Hungary's Roma minority. They emphasize that progress requires both internal change within the minority community and external support from the broader society: Singapore's Malay community improved educational outcomes through attitude shifts alongside government and societal support, a dual strategy that Hungary's Roma inclusion efforts have lacked. Focusing only on either minority self-improvement or majority-driven assistance is insufficient – a balanced approach was key to Singapore's success and is recommended for Hungary's Roma integration efforts.

Kálmán\cite{kalman2025} examines innovation policy through the lens of regulatory sandboxes for financial technology (FinTech) in the UK, Singapore, and Hungary. His comparative case study finds that sandbox programs markedly improve FinTech startups' access to capital, accelerate product development cycles, and facilitate their integration into the market. Singapore and the UK – both early adopters of sandboxes – serve as models with well-established, innovation-driven regulatory ecosystems, while Hungary's newer sandbox initiative illustrates the approach's potential in an emerging economy context. Kálmán concludes that by learning from Singapore's experience and tailoring sandbox design to local needs, Hungary can better balance innovation and oversight to foster FinTech growth.

\newpage
\section{Hungary's national AI strategy}
\label{sec:hungary-ai-strategy}

\subsection{Timeframe and responsible governmental body}
\label{subsec:timeframe-responsible-body}

\begin{itemize}
\item The AI Coalition, the agency established for the purpose of drafting the Strategy, was established in 2018 and drew the AI Action Plan in 2019, from which the Strategy was later formed from.
\item Hungary's AI Strategy has been published in 2020, and outlines vision and actions in the period of 2020 – 2030.
\item The Ministry of Innovation and Technology is responsible for the coordination and implementation of the strategy.
\end{itemize}

\subsection{Targeted sectors / areas to be supported}
\label{subsec:targeted-sectors}

The strategy's focus is on the following sectors:
\begin{itemize}
\item manufacturing
\item healthcare
\item agriculture
\item public administration
\item transportation
\item logistics and energy
\end{itemize}

They are categorized into three groups: foundation pillars; sector specific focus areas; and transformative programs

\subsection{Goals and actions}
\label{subsec:goals-actions}

\subsubsection{Foundation pillars}
\label{subsubsec:foundation-pillars}

The foundation pillars are focusing on creating a framework necessary to achieve developments in the field of AI. The below detailed \emph{Table~\ref{tab:foundation-pillars}} contains at least one from each of the categories of measures, selected to be the most relevant in terms of specific and measurable. In total, this part has 27 goals in 6 categories of measures.

\begin{table}[ht]
\centering
\caption{Foundation Pillars goals in Hungary's AI Strategy}
\label{tab:foundation-pillars}
\resizebox{\textwidth}{!}{%
\begin{tabular}{p{1cm}p{4cm}p{8cm}p{4cm}}
\toprule
\textbf{No.} & \textbf{Goal} & \textbf{Description} & \textbf{Deadline(s)} \\
\midrule
1. & Creation of a National Data Asset Agency 
\newline
\newline
\raggedright\emph{(Category: Setting the data economy in motion)} & An initiative for the adequate utilization of the public data assets. Support public bodies in possession of data assets, in keeping records of data inventories, in making them available for secondary use and in developing their business models. The organization and the associated public data inventory will be put in place on the basis of the results of projects already completed, projects under way and projects planned to be launched in this particular field. & \begin{minipage}[t]{4cm}
• Commissioned on: 01.10.2020\\
• Launch of public data portal (with usable data content): 31.03.2021\\
• Open access to at least 100 data sets: 31.01.2022\\
• Development of a public data inventory: 31.03.2021
\end{minipage} \\
\midrule
2. & Establishing the National Laboratory for Artificial Intelligence
\newline
\newline
\raggedright\emph{(Category: Research, development and innovation)} & Establish a consortium responsible for the cooperation in the academic sphere and for the conduct of harmonized AI research to meet the requirements of the market; establish and operate an organization system and a partnership framework ensuring that innovation sources are utilized and leveraged. & Founding date of the consortium: 31.12.2020 \\
\midrule
3. & Establishing the National Laboratory for Autonomous Vehicles
\newline
\newline
\raggedright\emph{(Category: same as 2.)} & Harmonize research and development activities associated with autonomous vehicles and channel these activities towards actual market demands. & Founding date of the consortium: 31.12.2020 \\
\midrule
4. & Support for the AI Accelerator Centre
\newline
\newline
\raggedright\emph{(Category: same as 2.}) & Build and develop an accelerator center for start-ups engaging in the development of or using the achievements of AI. & \begin{minipage}[t]{4cm}
• Date of tender invitation: 31.12.2020\\
• Date of publication of the funds: 31.07.2021
\end{minipage} \\
\midrule
5. & Establishing the Innovation Centre for Artificial Intelligence
\newline
\newline
\raggedright\emph{(Category: Incentivizing uptake)} & Build technology training research and infrastructure marketplaces; Create a team responsible for the support of AI implementation measures specifically aimed at small and medium enterprises (SMEs); Promote AI for SMEs & Launch the Innovation Centre on the base of the AI Coalition: 01.04.2021 \\
\midrule
6. & AI Challenge
\newline
\newline
\raggedright\emph{(Category: Education, competence development and societal preparedness)} & Train 100,000 people using internationally accredited online course material, Raise awareness of 1 million people via interactive exhibitions, a website and online professional contents & Start of challenge: 01.11.2020 \\
\midrule
7. & Enhancing supercomputer capacities
\newline
\newline
\raggedright\emph{(Category: Infrastructure development)} & As of 2022, the availability of 5 petaflops HPC capacity in Hungary & 31.03.2022 \\
\midrule
8. & Development of Hungarian testing environments
\newline
\newline
\raggedright\emph{(Category: Infrastructure development)} & Further develop testing environments, in particular connection to the European testing environment systems in the fields of self-driving vehicles, smart cities, agriculture and manufacturing; Encourage international cooperation opportunities and provision of the necessary conditions; Test network for Pre-Placement Readiness Demonstration PPRD and 5G purposes, test the government-owned smart settlement platform and infrastructure & \begin{minipage}[t]{4cm}
• Accreditation of ZalaZone as a European testing environment: 31.03.2021\\
• Accreditation of the agrarian model farms as a European testing environment: 31.03.2021
\end{minipage} \\
\midrule
9. & Creation of the Artificial Intelligence Regulation and Ethics Knowledge Centre
\newline
\newline
\raggedright\emph{(Category: Regulatory and Ethical Framework)} & Provide and coordinate a wide-ranging base of experts to support regulatory and ethics-related tasks. & 30.10.2020 \\
\bottomrule
\end{tabular}
}
\end{table}

\subsubsection{Sector specific focus areas}
\label{subsubsec:sector-specific}

Based on Hungary's economic structure and its strengths, a number of high priority sectors have been identified by the strategy, in which concerted efforts shall be made and concentrated. The below detailed \emph{Table~\ref{tab:sector-specific}} contains at least one from each of the categories of measures, selected to be the most relevant in terms of specific and measurable. In total, this part has 29 goals in 7 categories of measures.

\begin{table}[ht]
\centering
\caption{Sector specific goals in Hungary's AI Strategy}
\label{tab:sector-specific}
\resizebox{\textwidth}{!}{%
\begin{tabular}{p{1cm}p{4cm}p{8cm}p{3cm}}
\toprule
\textbf{No.} & \textbf{Goal} & \textbf{Description} & \textbf{Deadline} \\
\midrule
10. & Building of a manufacturing testing environment and integration into the EU
\newline
\newline
\raggedright\emph{(Category: Manufacturing)} & Build a testing environment capable of presenting manufacturing data analysis; Organize and exhibition of model applications; Provide grant for circular/green manufacturing & Setup of a test and model environment: 30.06.2021 \\
\midrule
11. & Development of AI applications in the fields of prevention, screening and diagnostics
\newline
\newline
\raggedright\emph{(Category: Healthcare)} & Develop AI applications in the field of imaging diagnostics; focused, preventive screenings based on the analysis of central files,; therapy and diagnostic decision support. Improve the efficiency of pharmaceutical research and support for in silico experiments; Develop medical technology equipment & Draft a detailed action plan: 31.03.2021 \\
\midrule
12. & Creation of a plot-based and data-driven farming consultancy service
\newline
\newline
\raggedright\emph{(Category: Agriculture)} & Based on the data originating from meteorological and spatial images or local farming, establish a central recommendation system and simple-to-implement, data-based advisory services & Launch of the advisory services: 30.06.2022 \\
\midrule
13. & Establishing an AI research cluster in public administration
\newline
\newline
\raggedright\emph{(Category: State Administration)} & Create a research cluster focusing on the use of AI applications by public bodies and central supply & 31.03.2022 \\
\midrule
14. & Extension of smart city concepts - traffic control, traffic management
\newline
\newline
\raggedright\emph{(Category: Transportation)} & Build technology training research and infrastructure marketplaces; Create a team responsible for the support of AI implementation measures specifically aimed at SMEs; Promotion of AI in for SMEs & Draw up the detailed project proposal: 31.03.2021 \\
\midrule
15. & Integrating smart meters, developing the conditions and requisites for data-driven processes
\newline
\newline
\raggedright\emph{(Category: Energy)} & Integrate smart meters, develope the conditions and requisites for data-driven processes & 2030 \\
\bottomrule
\end{tabular}
}
\end{table}

\subsubsection{Transformative programs}
\label{subsubsec:transformative-programs}

These goals aim to transform targeted sectors and encourage widespread AI use among citizens. The below detailed \emph{Table~\ref{tab:transformative-programs}} contains at least one from each of the categories of measures, selected to be the most relevant in terms of specific and measurable. In total, this part has 15 goals in 7 categories of measures.

\begin{table}[ht]
\centering
\caption{Transformative program goals in Hungary's AI Strategy}
\label{tab:transformative-programs}
\resizebox{\textwidth}{!}{%
\begin{tabular}{p{1cm}p{4cm}p{8cm}p{3cm}}
\toprule
\textbf{No.} & \textbf{Goal} & \textbf{Description} & \textbf{Deadline} \\
\midrule
16. & Building the infrastructure- and regulatory framework necessary for the operation of autonomous systems
\newline
\newline
\raggedright\emph{(Category: Transportation systems)} & Harmonize domestic and EU legislative environments and traffic police regulations. Develop the nationwide road network to contribute to the self-driving infrastructure. & Equip single-digit road networks with self-driving infrastructure by 2025 \\
\midrule
17. & Drafting of rating systems for healthcare data analytics applications
\newline
\newline
\raggedright\emph{(Category: Health consciousness in a digital world)} & Map digital healthcare applications; Develop rating criteria and monitoring processes; Communicate the rated service providers & 30.06.2021 \\
\midrule
18. & Preparing for the impacts of climate change and mitigating its adverse effects
\newline
\newline
\raggedright\emph{(Category: Climate-driven agriculture)} & Develop and apply AI-based, optimization solutions in terms of plant production and stock farming & Draw up the detailed project proposal: 31.12.2020 \\
\midrule
19. & Development of the data wallet technology model
\newline
\newline
\raggedright\emph{(Category: Data wallet and personalized services)} & Develop software to support citizens in making statements regarding the use, sale or disclosure of their data to third parties in a one-stop-shop system. Integrate Hungarian patents to allow fully anonymous accesses. & \begin{minipage}[t]{3cm}
Prepare the first demo model: 31.12.2020\\
Build the model providing full anonymity: 28.02.2022
\end{minipage} \\
\midrule
20. & AI-supported career advisory system
\newline
\newline
\raggedright\emph{(Category: AI-supported development of personal competences)} & Develop the necessary files, expert activities, ensuring interoperability of file systems; Develop a personalized training recommendation service that is tailored to individual life objectives and based on public education, vocational training, tertiary vocational training and adult education offers, along with the labor market results attained by their completion & Draw up the detailed project proposal: 30.06.2021 \\
\midrule
21. & Support for implementation in customer services
\newline
\newline
\raggedright\emph{(Category: Automated administrative procedures in Hungarian)} & Expand to government administrative matters; Tenders for the development of AI-supported customer services and for the building the 1818 Government Client Hotline and for cooperation regarding infocommunications & Launch of wide-scale promotion: 31.12.2022 \\
\midrule
22. & Implementation of smartgrid technologies
\newline
\newline
\raggedright\emph{(Category: Energy networks focused on renewable sources of energy)} & Build upon smart meters and implement smartgrid technologies and wide-scale forecast systems; Facilitate the creation of a more accurate production timeline for weather-dependent renewable energy sources and the operation of the energy network relying on it & Implement a new network development and network connection regulation by 2025 to support an efficient system integration of the production of renewable energy \\
\bottomrule
\end{tabular}
}
\end{table}
\FloatBarrier

\subsection{Assessment of goals and actions}
\label{subsec:assessment-goals}

\begin{table}[ht]
\centering
\caption{Assessment of Hungary's AI Strategy: goals 1-6 (as of February 2025)}
\label{tab:assessment-goals-1-6}
\footnotesize
\begin{tabular}{>{\raggedright\arraybackslash}p{0.8cm}>{\raggedright\arraybackslash}p{2.8cm}>{\arraybackslash}p{11.5cm}}
\toprule
\textbf{No.} & \textbf{Goal} & \textbf{Assessment (as of 02.2025)} \\
\midrule

1. & Creation of a National Data Asset Agency & \href{https://navu.hu/about}{\textcolor{blue}{NAVÜ}} (Nemzeti Adatvagyon Ügynökség) was announced by the government in \href{https://kormany.hu/hirek/megalakult-a-nemzeti-adatvagyon-ugynokseg}{\textcolor{blue}{October 2020}} , and registered as a company in \href{https://www.ceginformacio.hu/cr9312108599}{\textcolor{blue}{August 2021}} with a starting budget of HUF 50 million (\textasciitilde EUR 125,000). It serves as a central body in Hungary whose main task is to facilitate the use of the national data assets held by public bodies, including public data, personal data and protected data.

2,283 datasets are currently available for download, in 11 categories: General Public Services; Health; Economic Affairs; Environment; Public Order and Security; Housing and Urban Development; Education; Leisure, Culture and Religious Affairs; Social Protection; Defence \\
\midrule
2. & Establishing the National Laboratory for Artificial Intelligence & The date of the establishment of the consortium is unknown, but the corresponding tender for the establishment and complex development of a National Laboratory was announced in \href{https://nkfih.gov.hu/palyazoknak/aktualis-felhivasok/egyeb-tamogatas/nemzeti-laboratorium-letrehozasa-rrf-231-2021/palyazati-felhivas}{\textcolor{blue}{December 2021}}, with a budget of HUF 72 billion (\textasciitilde EUR 180,000,000), of which HUF 9 billion (\textasciitilde EUR 23,000,000) was \href{https://milab.tk.hu/hirek/2022/11/a-projekt-alapadatai}{\textcolor{blue}{allocated}} for establishment and complex development of the \href{https://mi.nemzetilabor.hu/about-us}{\textcolor{blue}{National Laboratory for Artificial Intelligence (MILAB)}}.

MILAB operates as a consortium of Hungarian scientific institutions, coordinated by the Institute for Computer Science and Control (SZTAKI). MILAB has 6 areas of search: Theoretical Foundations, Machine Vision, Language Technology, Medical AI and Diagnostics, Sensors, IoT and Telecommunications, Security and Data Protection. According to their \href{https://mi.nemzetilabor.hu/publications?page=0}{\textcolor{blue}{website}}, 41 publications have been published by MILAB, but assuming that the latest one is dated 2021, the website might not have been updated since, and the number of current publications may be much higher. \\
\midrule
3. & Establishing the National Laboratory for Autonomous Vehicles & The date of the establishment of the consortium is unknown, but the corresponding tender for the establishment of the \href{https://autonom.nemzetilabor.hu}{\textcolor{blue}{Autonomous Vehicles National Laboratory}} has a funding period starting from \href{https://www.nemzetilaborok.nkfih.gov.hu/digital-transformation/files-to-download-240826/national-laboratory-for-240826}{\textcolor{blue}{April 2022}}, with an allocated budget of HUF 6 billion (\textasciitilde EUR 16,000,000).

The consortium is led by the Institute for Computer Science and Control (SZTAKI), with the Budapest University of Technology and Economics and the Széchenyi University as partners. Research is done in the field of autonomous road vehicles, autonomous aerial vehicles, autonomous robotics and production systems. According to their\href{https://autonom.nemzetilabor.hu/publications?page=0}{\textcolor{blue}{website}}, 229 publications have been published. \\
\midrule
4. & Support for the AI Accelerator Centre & The publication of the funds was done in \href{https://neum.hu/328-2/}{\textcolor{blue}{January 2021}}, which was earlier than the planned deadline of July 2021. HUF 1 billion (\textasciitilde EUR 2,500,000) was allocated for the establishment of two centers in \href{https://ai-hungary.com/akcelerator/hu/akcelerator-kozpont/adatgazdasag-akcelerator-kozpont-debrecen}{\textcolor{blue}{Debrecen}} and \href{https://ai-hungary.com/akcelerator/hu/akcelerator-kozpont/mesterseges-intelligencia-akcelerator-kozpont-zalaegerszeg}{\textcolor{blue}{Zalaegerszeg}}. Apart from the centers, a \href{https://ai-hungary.com/akcelerator}{\textcolor{blue}{website}} was created to help SME-s and startups in adopting AI, where learning materials, best practices, consultation opportunities and workshops can be found. No information could be found on the internet how well the two centers are helping the target groups, neither is it possible, currently, to have an appointment for the consultation service on the website. \\
\midrule
5. & Establishing the Innovation Centre for Artificial Intelligence & No information is available online about how this differs from Goal No. 4 or whether this goal has separate results, therefore, it is likely that the two were merged. \\
\midrule
6. & AI Challenge & The corresponding website where the AI Challenge was located was launched in \href{https://whoisfreaks.com/tools/user/whois/history/lookup/ai-hungary.com}{\textcolor{blue}{May 2020}}. On the \href{https://ai-hungary.com/hu/tartalom/mi-akademia/ertsd-meg}{\textcolor{blue}{website}} , apart from the challenge, podcasts and videos can be found with the goal of spreading awareness about the potential of AI. No information could be found on the internet about whether the milestone of 100,000 people have completed the survey. The domain (\href{https://www.ai-hungary.com/}{\textcolor{blue}{https://www.ai-hungary.com/}}) no longer belongs to the responsible government agency and was reserved by a gambling company. This at the time created \href{https://telex.hu/techtud/2025/01/16/kaszinoreklam-fut-egy-volt-kormanyzati-honlapon-ai-hungary-com-digitalis-jolet-kft}{\textcolor{blue}{press coverage}} on the case, as the new owner kept the old design and logos to promote gambling services. The separate domain for the challenge (\href{https://mikihivas.hu/}{\textcolor{blue}{https://mikihivas.hu/}}) does not function anymore as both domains were purchased only until July 2024. The website was reallocated \href{https://mik.neum.hu}{\textcolor{blue}{here}}, where the \href{https://account.nexiuslearning.com/login?returnUrl=%2FWSFederation%2FIssue%3Fwtrealm%3Dhttps%253A%252F%252Fhome.nexiuslearning.com%252F%26wa%3Dwsignin1.0%26wreply%3Dhttps%253A%252F%252Fhome.nexiuslearning.com%252Fsignin-wsfed%26wctx%3DCfDJ8JfsSc_ybbtJj8E1YtWdhGdCcM5Fq5TxvxqhBf3pEImbpOoryTPdbyRBfcWxrgrubdCzs6vtAIWgf-VmVXpDhl8V0ccWjWGJh3Zn46_gQjRpJAHbEqbguQH89l1CyPoZqlJR0vCTB6J9gQgWmq8XKZ4sUW7r-44UOwkSQPM7Q9739FCfOLqURrf4CAEKmuW0NKx-RbPzuNNMVLYZm9gPCYeHwIsF2RYt2StoLHjgDix8rSHSyoCyOIiNDr1nqSKc4s20cn_JIkEZjwWZ1KKIotRjjHjdHR3TYVaKJVKO4xSL&wtrealm=https:%2F%2Fhome.nexiuslearning.com%2F&wa=wsignin1.0&wreply=https:%2F%2Fhome.nexiuslearning.com%2Fsignin-wsfed&wctx=CfDJ8JfsSc_ybbtJj8E1YtWdhGdCcM5Fq5TxvxqhBf3pEImbpOoryTPdbyRBfcWxrgrubdCzs6vtAIWgf-VmVXpDhl8V0ccWjWGJh3Zn46_gQjRpJAHbEqbguQH89l1CyPoZqlJR0vCTB6J9gQgWmq8XKZ4sUW7r-44UOwkSQPM7Q9739FCfOLqURrf4CAEKmuW0NKx-RbPzuNNMVLYZm9gPCYeHwIsF2RYt2StoLHjgDix8rSHSyoCyOIiNDr1nqSKc4s20cn_JIkEZjwWZ1KKIotRjjHjdHR3TYVaKJVKO4xSL}{\textcolor{blue}{challenge}} is also available after signing up. \\
\bottomrule
\end{tabular}
\end{table}

\begin{table}[ht]
\centering
\caption{Assessment of Hungary's AI Strategy: goals 7-10 (as of February 2025)}
\label{tab:assessment-goals-7-10}
\footnotesize
\begin{tabular}{>{\raggedright\arraybackslash}p{0.8cm}>{\raggedright\arraybackslash}p{2.8cm}>{\arraybackslash}p{11.5cm}}
\toprule
\textbf{No.} & \textbf{Goal} & \textbf{Assessment (as of 02.2025)} \\
\midrule
7. & Enhancing supercomputer capacities & The supercomputer named "Komodor" was purchased for nearly HUF 7 billion (\textasciitilde EUR 17,600,000) and placed in Debrecen. It began operating by the end of \href{https://hirek.unideb.hu/en/komondor-hungarys-most-powerful-supercomputer-awakening}{\textcolor{blue}{2022}}. A corresponding \href{https://hpc.kifu.hu/hu/komondor.html}{\textcolor{blue}{webpage}} was created with the goal of enabling SME-s to use the 5-petaflop computer.

In June 2022, the government \href{https://kormany.hu/hirek/nyertes-palyazatnak-koszonhetoen-epul-uj-magyar-szuperszamitogep-csillebercen-a-jelenlegi-kapacitas-negyvenszereset-tudja-majd-levente}{\textcolor{blue}{announced}} a new computer named ``Levente'' to be built with the computing capacity of 20 petaflops, four times greater than Komodor's. The EU funding for the project amounts to 7 billion HUF (\textasciitilde EUR 17,600,000). No information could be found on how much the local government contributes. \\
\midrule

8. & Development of Hungarian testing environments & Both Zalazone and agrarian model farms Mezőhegyes National Stud farm received numerous investments from the government and the EU for development during the period between 2020 – 2025. For example, the Széchenyi István University \href{https://admissions.sze.hu/sze-continues-to-expand-its-zalazone-development-base}{\textcolor{blue}{secured}} HUF 4 billion (\textasciitilde EUR 11,400,000) for infrastructure development at Zalazone; Mezőhegyes National Stud farm received Hungary's largest irrigation \href{https://trademagazin.hu/en/elkeszult-magyarorszag-legnagyobb-ontozesfejlesztesi-beruhazasa-mezohegyesen/}{\textcolor{blue}{development}} project in 2022 worth HUF 10.5 billion (\textasciitilde EUR 29,000,000).

\begin{itemize}
\item Zalazone: No official announcement or confirmation was made about the accreditation of Zalazone as European testing environment. The European Commission's \href{https://digital-strategy.ec.europa.eu/en/faqs/testing-and-experimentation-facilities-tefs-questions-and-answers}{\textcolor{blue}{Testing and Experimentation Facilities}} (TEFs) – which provide formal ``European testing'' designations – launched in \href{https://www.fbk.eu/en/press-releases/europe-launches-four-large-scale-ai-test-facilities/}{\textcolor{blue}{mid-2023}}. TEF is a collaboration of 33 partners across 11 countries. In the ``Smart Cities and Communities'' category, which is run under the name and website of \href{https://citcomtef.eu/}{\textcolor{blue}{https://citcomtef.eu/}}, focuses on domains like power, connectivity, and mobility (e.g. self-driving cars), but Hungary and Zalazone \href{https://citcom.ai/services}{\textcolor{blue}{cannot}} be found concerning the listed testing sites.

\item Hungarian agrarian model farms: No official announcement or confirmation was made about the accreditation of any Hungarian model farms as European testing environment. TEF's relevant branch is agrifoodTEF, also launched in mid-2023. Even Hungary's flagship digital farm (the Mezőhegyes National Stud Farm's smart agriculture center) is not part of the project as of agrifoodTEF's \href{https://www.agrifoodtef.eu/catalogue-of-services?search_api_fulltext=}{\textcolor{blue}{official catalogue}}.
\end{itemize}\\
\midrule
9. & Creation of the Artificial Intelligence Regulation and Ethics Knowledge Centre & A department named Centre for Artificial Intelligence Regulation and Ethical Knowledge (MISZET) was \href{https://www.elte.hu/content/a-mesterseges-intelligencia-emberkozpontu-korszakanak-kuszoben-allunk.t.30804}{\textcolor{blue}{established}} at Eötvös Loránd Tudományegyetem (ELTE) with the goal of ``playing a key role in coordinating the legal regulation of AI and developing educational programs in Hungary''. No website, intellectual product or publication concerning MISZET could be found on the internet. \\
\midrule
10. & Building of a manufacturing testing environment and integration into the EU & No information is available online about how this differs from Goal No. 8 or whether this goal has separate results, therefore, it is likely that the two were merged. \\
\bottomrule
\end{tabular}
\end{table}

\begin{table}[ht]
\centering
\caption{Assessment of Hungary's AI Strategy: goals 11-13 (as of February 2025)}
\label{tab:assessment-goals-11-13}
\footnotesize
\begin{tabular}{>{\raggedright\arraybackslash}p{0.8cm}>{\raggedright\arraybackslash}p{2.8cm}>{\arraybackslash}p{11.5cm}}
\toprule
\textbf{No.} & \textbf{Goal} & \textbf{Assessment (as of 02.2025)} \\
\midrule
11. & Development of AI applications in the fields of prevention, screening and diagnostics & No detailed action plan could be found on the internet, but the following developments occurred as cooperation between the public and the private sector. By 2024, officials were integrating various digital tools into public health programs (like the national E-Health app \href{https://www.eeszt.gov.hu/hu/eeszt-mobilalkalmazas}{\textcolor{blue}{``EgészségAblak''}} for e-invitations to doctors. See details as part of Goal number 17) to improve participation and support diagnostics, from which the most important ones are summarized:

\begin{itemize}
\item Oncompass Medicine, a Budapest-based health-tech company, developed an AI-driven software platform that helps oncologists select the optimal targeted therapy based on a patient's molecular profile. This precision oncology decision support system – the Oncompass ``Calculator'' – interprets tumor genetic data to match patients with the best personalized treatment options. Oncompass's innovation earned the 2021 Future Unicorn Award, a pan-European \href{https://monitor-industrial-ecosystems.ec.europa.eu/news/hungarian-ai-based-healthcare-company-winner-future-unicorn-award-2021}{\textcolor{blue}{recognition}} for high-potential tech scale-ups. Since then, the software has been implemented in Hungary's oncology practice: for example, in onco-hematology it was integrated with the National Healthcare Service Center's systems. A 2019–2020 project led by the National Institute of Oncology and Oncompass created a nationwide precision medicine decision support \href{https://pubmed.ncbi.nlm.nih.gov/31821382}{\textcolor{blue}{framework}}, combining a ``OncoGenomic'' data management program with the self-learning Oncompass AI tool. These tools feed into a dynamic national cancer registry and exchange data via the Elektronikus Egészségügyi Szolgáltatási Tér (EESZT, Electronic Health Service Space, read more from Goal 17) e-health cloud, ensuring that patients across Hungary have equal access to cutting-edge molecular diagnostics and therapies.
\item Beyond oncology, AI is being used to harness unstructured clinical data for both research and care. A two-year R\&D project (completed in 2023) between the University of Debrecen and GE HealthCare – funded by the National Research, Development and Innovation Fund – developed an \href{https://hungarytoday.hu/hungarian-innovation-for-better-analysis-of-medical-findings-using-ai/}{\textcolor{blue}{AI solution}} to better analyze medical findings. With a budget of HUF 1.5 billion (\textasciitilde EUR 3,900,000), half from public grant), this project created natural language processing algorithms for Hungarian medical text, integrated with diagnostic imaging data. The result is a software tool that can search and interpret radiology images and doctors' written notes simultaneously.
\end{itemize} \\
\midrule
12. & Creation of a plot-based and data-driven farming consultancy service & Specific information confirming the official launch of these advisory services could not be found publicly. However, there has been development in state owned agencies:

\begin{itemize}
\item Hungarian Meteorological Service (OMSZ) \href{https://asr.copernicus.org/articles/20/9/2023/}{\textcolor{blue}{offers}} a variety of agrometeorological services to farmers through a dedicated subpage. These services include observations, specialized forecasts, drought monitoring, and regular analyses of crop conditions, all aimed at supporting agricultural decision-making. Experts also provide advice via telephone, media interviews, and publications in agricultural magazines.
\item Research indicates that Hungarian agribusinesses are \href{https://www.mdpi.com/2071-1050/13/18/10269}{\textcolor{blue}{increasingly}} adopting Climate-Smart Agriculture CSA solutions, which involve the use of digital tools and data-driven practices to enhance sustainability and productivity.
\item Research shows that Hungary has made \href{https://www.researchgate.net/publication/353747802_Earth_observation_and_geospatial_big_data_management_and_engagement_of_stakeholders_in_Hungary_to_support_the_SDGs}{\textcolor{blue}{advancements}} in earth observation and geospatial data management, supporting sustainable development goals.
\end{itemize}

Concerning consultancy services, recently, the \href{https://ec.europa.eu/regional_policy/en/newsroom/news/2020/06/06-10-2020-stories-from-the-regions-new-technologies-in-agriculture-in-hungary}{\textcolor{blue}{ three-way consortium}} of the agricultural service provider KITE, the University of Debrecen and Balogh-Farm introduced new precision technology in agriculture. \\
\midrule
13. & Establishing an AI research cluster in public administration & As of the time of writing this article, there was no announcement of the establishment of a standalone AI research cluster in the field of public administration. In practice, however, Hungary built up AI R\&D networks and programs that fulfilled to some extent this goal. Instead of a single new entity launched on that date, the government probably leveraged existing initiatives – notably the National Artificial Intelligence Laboratory and the AI Coalition – to serve as the de facto research cluster for public-sector. Although there might be several implemented AI projects in the public sector (e.g., in the field of security and law enforcement, document processing), there are no public information is available about the existence of such systems. \\
\bottomrule
\end{tabular}
\end{table}

\begin{table}[ht]
\centering
\caption{Assessment of Hungary's AI Strategy: goals 14-16 (as of February 2025)}
\label{tab:assessment-goals-14-16}
\footnotesize
\begin{tabular}{>{\raggedright\arraybackslash}p{0.8cm}>{\raggedright\arraybackslash}p{2.5cm}>{\arraybackslash}p{12cm}}
\toprule
\textbf{No.} & \textbf{Goal} & \textbf{Assessment (as of 02.2025)} \\
\midrule
14. & Extension of smart city concepts - traffic control, traffic management & The details of the plan for launching an AI Innovation Centre can be found at Goal 4.

Since 2021, several pilot projects and research initiatives have been launched, with the goal of nationwide deployment of smart city concepts, traffic control and management. However, as of the time of writing this article, full implementation in Hungarian cities is not yet done. The most important highlights are summarized:
\begin{itemize}
\item Szombathely, as part of an EU \href{https://raketa.hu/ket-magyar-varos-is-bejutott-tavaly-az-eu-100-intelligens-varos-kihivas-nevu-programjaba}{\textcolor{blue}{``Intelligent Cities Challenge''}} program, has plans to develop intelligent parking and traffic control systems that use data analytics and AI.
\item On the R\&D side, a four-year project called \href{https://magyarepitok.hu/iparagi-hirek/2024/04/elkeszult-korunk-legpontosabb-forgalomiranyitasi-rendszere}{\textcolor{blue}{iTrafficTools}} (started April 2020) was successfully completed in March 2024 by engineering firm Főmterv and BME University. This project created one of the first dynamic, AI-supported traffic control models, capable of adjusting traffic light strategies in real time based on live data – a significant milestone toward smarter traffic management.
\item Budapest's Medianets Lab (BME) is co-developing an AI-based traffic light control system, \href{https://www.itsinternational.com/its4/its5/its8/news/new-ai-traffic-project-developed-hungary-turkey-and-japan}{\textcolor{blue}{TRALICO}}, with partners in Turkey and Japan. This system uses deep learning to forecast congestion and optimize signals and is set to be tested on the streets of Istanbul.
\end{itemize} \\
\midrule
15. & Integrating smart meters, developing the conditions and requisites for data-driven processes & Currently, Hungary is still in the \href{https://balkangreenenergynews.com/central-eastern-europe-severely-lagging-in-smart-meters-rollout/}{\textcolor{blue}{early stages}} of smart meter integration. As of early 2024, Hungary has installed approximately 800,000 smart meters, with E.ON Hungária Group responsible for two-thirds of these installations. The government aims to \href{https://hungarytoday.hu/smart-meters-to-contribute-to-the-countrys-optimal-energy-consumption/}{\textcolor{blue}{increase}} this number to around 4.7 million units by 2030, which shows that the current progress lags behind the expected. See more detailed information at Goal 22. \\
\midrule

16. & Building the infrastructure- and regulatory framework necessary for the operation of autonomous systems & The evaluation of the goal of equipping all single-digit national roads (the 8 most important highways) with ``self-driving'' infrastructure by 2025 is hard to evaluate, since no precise definition for such infrastructure exists in the document, and no such accepted and widespread standards exists. There have been no announcements of fully outfitting every major route yet. However, various pilot projects were started, the most notable being:
\begin{itemize}
\item Short-range V2X transmitters were \href{https://internet.kozut.hu/2019/04/23/a-magyar-kozut-a-szomszedos-orszagok-kozutkezeloivel-kozosen-keszul-az-onvezeto-autozasra}{\textcolor{blue}{installed}} on the M0 ring road around Budapest as early as 2019, laying groundwork for connected and automated driving corridors.
\item A multi-day ``intelligent road'' test on the M86 highway (Csorna bypass) in June 2020 was \href{https://24.hu/tech/2020/06/25/m86-intelligens-infrastruktura-onvezeto-jarmuvek/}{\textcolor{blue}{organized}}, led by the Budapest University of Technology and Economics with partners Knorr-Bremse, Hungarian Public Roads (Magyar Közút), Ericsson, Magyar Telekom, and Austria's Virtual Vehicle research institute. During this pilot, a closed 7 km highway section was \href{https://www.hwsw.hu/hirek/61957/onvezeto-auto-intelligens-infrastruktura-teszt-lidar-radar-5g-szenzor-iot-kozut.html}{\textcolor{blue}{equipped}} with dozens of sensors and 5G connectivity and populated with sensor-loaded cars and a truck to simulate autonomous driving scenarios
\item In mid-2024, Hungarian Public Roads (Magyar Közút) awarded a \href{https://magyarepitok.hu/mi-epul/2024/07/knight-rider-a-magyar-utakon-kialakitjak-a-kijelolt-autopalyaszakaszon-az-onvezeto-jarmuvek-mukodeset-tamogato-szenzorszigeteket}{\textcolor{blue}{contract}} to Sagemcom to install ``sensor islands'' along the M1–M7 highway corridor near Budapest. Each sensor hub includes multiple cameras (wide and narrow-angle per lane), a fisheye camera, thermal imaging, radar units, and both long-range and short-range LiDAR sensors. These roadside units will stream real-time traffic and environmental data to a new central system, assisting autonomous vehicle navigation and testing on the highway. According to the tender, the sensor islands and central system are operational by autumn 2024.
\end{itemize} \\
\bottomrule
\end{tabular}
\end{table}

\begin{table}[ht]
\centering
\caption{Assessment of Hungary's AI Strategy: goals 17-18 (as of February 2025)}
\label{tab:assessment-goals-17-18}
\footnotesize
\begin{tabular}{>{\raggedright\arraybackslash}p{0.8cm}>{\raggedright\arraybackslash}p{2.5cm}>{\arraybackslash}p{12cm}}
\toprule
\textbf{No.} & \textbf{Goal} & \textbf{Assessment (as of 02.2025)} \\
\midrule
17. & Drafting of rating systems for healthcare data analytics applications & No evidence could be found that a certification or rating portal for health apps (similar to the NHS Digital Apps Library in the UK or Germany's DiGA directory) had been launched by the target date of June 2021, neither as of the time of writing this article.

However, the government took several steps towards this goal with the improvement and digitalization of national healthcare systems. The most notable were:
\begin{itemize}
\item The government launched a \href{https://okfo.gov.hu/egeszsegugyi-fejlesztesek/hazai-fejlesztesek/2014-2020-fejlesztesi-idoszak/lezarult-fejlesztesek/efop-1-9-6-16-2017-00001}{\textcolor{blue}{tender}} for the digitalization of the health sector enabling HUF 65 billion (\textasciitilde EUR 160,000,000), with five main goals:
\begin{itemize}
\item Central service development and further development of the EESZT (detailed in a separate point)
\item Establishment of a central teleconsultation and telemedicine framework system, implementation of a telemedicine pilot
\item Utilization of sectoral data assets
\item Development of citizens' e-health competencies, provision of business-oriented utilization functions and verified electronic information
\item Development of sectoral IT infrastructure
\end{itemize}
\item EESZT: An infrastructure enabling secure integration of patient-generated health data from smart devices and health applications. By 2022, a new \href{https://e-egeszsegugy.gov.hu/mobilgateway}{\textcolor{blue}{``Mobil Gateway''}} interface was being developed to allow approved mobile health apps to connect with EESZT's public services securely. This technical framework lets apps upload data (e.g. from wearables, home sensors) into patients' electronic records. It creates a form of monitoring process, as data from digital health tools can now be channeled into the healthcare system for oversight. There is no provided catalogue of certified apps or any star-ratings.
\item In 2022, the government launched \href{https://egeszsegvonal.gov.hu/}{\textcolor{blue}{``Egészségvonal''}} (Health Line), a public health information portal aimed at improving health awareness. Among its content, Egészségvonal includes sections on ``Digital Health'' and ``Smart devices and applications'', educating citizens on using health apps and gadgets responsibly.
\end{itemize} \\
\midrule
18. & Preparing for the impacts of climate change and mitigating its adverse effects & No official announcement could be found about a detailed action plan on the internet as of the time of writing this article.

From 2021 onward, the government made agrometeorological data free for farmers (via a website called \href{https://www.met.hu/idojaras/agrometeorologia/}{\textcolor{blue}{HungaroMet}}, run by the Hungarian Meteorological Service). This open data access enables AI-driven decision support (e.g. smarter irrigation scheduling and weather-risk predictions), directly supporting the climate adaptation goal.

The government also outlined (back in 2019) the \href{https://neum.hu/das/}{\textcolor{blue}{Digitális Agrár Stratégia}} (Digital Agriculture Strategy) for preparing Hungary's agri-sector for advanced technologies. Apart from that, The National Agriculture Chamber (NAK) created a TechLab incubator in 2019 to accelerate agritech startups. By 2021, over 50 agrár-tech startups had been nurtured. \\
\bottomrule
\end{tabular}
\end{table}

\begin{table}[ht]
\centering
\caption{Assessment of Hungary's AI Strategy: goal 19 (as of February 2025)}
\label{tab:assessment-goals-19}
\footnotesize
\begin{tabular}{>{\raggedright\arraybackslash}p{0.8cm}>{\raggedright\arraybackslash}p{2.5cm}>{\arraybackslash}p{12cm}}
\toprule
\textbf{No.} & \textbf{Goal} & \textbf{Assessment (as of 02.2025)} \\
\midrule
19. & Development of the data wallet technology model & The goal was to create a one-stop-shop software tool through which individuals could manage consent for the use, sharing, or even sale of their personal data, while remaining in full control. This data wallet concept was envisioned as a transformative project to give citizens ``full control over their data'', letting them decide who can access their data.

\begin{itemize}
\item GINOP 3.2.8-20 EU and its subproject \href{https://neum.hu/en/328-2/}{\textcolor{blue}{GINOP-3.2.8-20-2020-00001}} provided the financial background for the realization of the above-mentioned goal, with a sum of HUF 1 billion (\textasciitilde EUR 2,500,000)
\item With the establishment of NAVÜ (detailed as the 2nd goal), a framework and legislation (Act XCI of 2021, effective June 2021) was created.
\item In 2021, the Neumann Nonprofit Közhasznú Kft. (a state-owned digital innovation company) launched the \href{https://neum.hu/en/natuk/}{\textcolor{blue}{National Data Economy Knowledge Center}}
 (Nemzeti Adatgazdasági Tudásközpont, NATUK). The Knowledge Center's chief mission is supporting the AI Strategy's implementation – notably \href{https://ivsz.hu/hirek/diktaljuk-mi-a-tempot-az-adattarcakkal-interju-tarcsi-adammal/}{\textcolor{blue}{bootstrapping}} the domestic data economy and ecosystem. It focuses on the ``data'' side of AI: developing the data wallet, advising on data economy regulations, building economic models for data valuation, and helping small businesses leverage data.
\end{itemize}

After the groundwork was completed by 2022, the project's focus shifted toward integration into the development of a national digital identity and services platform. This platform allows citizens to store all forms of identification—such as driving license and ID card—in one place, with the goal of achieving nationwide acceptance. The project's aim is to fulfill the EU eIDAS 2.0 regulation for a national e-ID wallet and to unify e-government services.

This initiative has now a much larger funding background, as it is now incorporated into larger scale program called \href{https://www.palyazat.gov.hu/hirek/dimop-plusz-uj-lendulet-magyarorszag-digitalis-fejlodesehez}{\textcolor{blue}{DIMOP Plusz}}. launched in 2022 with 80\% funding from the European Union, with a total budget of HUF 732 billion (\textasciitilde EUR 1,800,000,000), available for its implementation until 2029. The project called DIMOP Plusz is linked to one of the programs from the Széchenyi 2020 development period, the Public Administration and Public Services Development Operational Program (KÖFOP), which contributes to the modernization of Hungary's electronic public administration systems and the transition of certain services to the digital space. Since this development, the following progress has been made:
\begin{itemize}
\item The Digital Personal Data Wallet service, which officially launched in September 2024, enables users to store personal data certificates and manage consent for data sharing.
\item As of early 2025, the ecosystem is being prepared for wider adoption. The DÁP-sandbox is allowing companies to integrate and test their systems with the data wallet API.
\end{itemize}

In conclusion, the data wallet project laid the foundation for citizen-controlled data sharing through legal, institutional, and financial support. It evolved or merged into a broader national digital identity platform aligned with EU standards, enabling centralized access to official documents. Since companies did not yet integrate the service into their own application, the use of the Digital Personal Data Wallet is not yet widespread and therefore limited in its capabilities. \\
\bottomrule
\end{tabular}
\end{table}

\begin{table}[ht]
\centering
\caption{Assessment of Hungary's AI Strategy: goals 20-21 (as of February 2025)}
\label{tab:assessment-goals-20-21}
\footnotesize
\begin{tabular}{>{\raggedright\arraybackslash}p{0.8cm}>{\raggedright\arraybackslash}p{2.5cm}>{\arraybackslash}p{12cm}}
\toprule
\textbf{No.} & \textbf{Goal} & \textbf{Assessment (as of 02.2025)} \\
\midrule
20. & AI-supported career advisory system & As of the time of writing this article, such AI-supported career advisory system has not been launched as a standalone public website or app. However, there have been different projects related to career advisory:
\begin{itemize}
\item \href{https://ikk.hu/}{\textcolor{blue}{National Career Orientation Portal}} (Nemzeti Pályaorientációs Portál): A one-stop online portal for career guidance maintained by the education and labor authorities. This portal provides information on professions, training opportunities, job search tips, and self-assessment questionnaires for various age groups, but there are no signs that it has any kind of AI-related function.
\item {Pályaorientációs Mérő- és Támogatóeszköz}: A Career Orientation Assessment and Support Tool developed by the ``Oktatási Hivatal'' (Educational Office). This tool was created under a European Social Fund project (\href{https://www.palyazat.gov.hu/programok/szechenyi-2020/efop/efop-3213-17/alapadatok}{\textcolor{blue}{EFOP-3.2.13-17}}) with HUF 1 billion funding (\textasciitilde EUR 250,000,000). POM is essentially an online library of self-evaluation questionnaires for school students to gauge their interests, strengths, and STEM inclinations. Based on the answers, the system suggests suitable vocational or educational directions. The development was completed by September 2021, and thousands of teachers were \href{https://www.oktatas.hu/kozneveles/palyaorientacio/mero_tamogatoeszkoz_POM}{\textcolor{blue}{trained}} to use it in schools. POM is accessible to all graduating primary and secondary students. Although it uses predefined algorithms (not machine-learning) to match students with career profiles, it represents a personalized digital guidance service.
\item {Career Intelligence Project (Karrier 4.0 / Career Intelligence – Erasmus+)}: A European Union funded pilot focusing on AI in career planning for youth. Under the leadership of Ruhr-Universität Bochum (Germany) and with Hungarian \href{https://pbkik.hu/2023/07/18/hirek/karriertervezes-a-mesterseges-intelligencia-tamogatasaval-bemutatkozott-a-career-intelligence-projekt/#:~:text=A%20%E2%80%9EKarrier%204,p%C3%A1lyaorient%C3%A1ci%C3%B3ban%20hivatott%20t%C3%A1mogatni%20a%20fiatalokat}{\textcolor{blue}{participation}} (Pécs-Baranya Chamber of Commerce), this project created an online learning and career planning platform enriched with AI tools. While this is an international pilot (as part of the EU's Erasmus+ program) rather than a national service, it can serve as an example.
\end{itemize} \\
\midrule
21. & Support for implementation in customer services & The state IT agency NISZ (National Infocommunications Service Zrt.) was tasked with the 1818 hotline modernization, developing the \href{https://nisz.hu/sajtoszoba/egyre-hatekonyabban-mukodik-a-mesterseges-intellig-d158}{\textcolor{blue}{MIA}} chatbot in-house (through its software arm IdomSoft). A chatbot named MIA was launched in May 2021 as a virtual assistant for administrative customer service. MIA is a hybrid AI chatbot deployed on the 1818 Government Hotline's web chat channel, combining machine learning with human oversight. It assists both citizens and call center agents by suggesting answers and learning from confirmed responses. Since launch, MIA has handled nearly \href{https://nisz.hu/sajtoszoba/egyre-hatekonyabban-mukodik-a-mesterseges-intellig-d158}{\textcolor{blue}{1,000,000 inquiries}}, with 95\% of answers supported by AI and 38\% of 2022 queries resolved fully automatically without human intervention.

To expand AI-supported administration beyond the call center, the government introduced self-service kiosks (MIA Pontok) in physical client service centers (government ``Kormányablak'' offices). Pilots started in 2021–2022 with kiosks in a \href{https://www.parlament.hu/documents/10181/39233854/Infojegyzet_2021_73_mesterseges_intelligencia_a_kozigazgatasban.pdf/#:~:text=A%20KIOSK%20projekt%20keret%C3%A9ben%20k%C3%ADs%C3%A9rleti,darabot%20fog%02nak%20bel%C5%91l%C3%BCk%20orsz%C3%A1gszerte%20a}{\textcolor{blue}{few offices}}. Each kiosk is an intelligent terminal equipped with a touch-screen, camera, microphone, fingerprint reader, ID scanner, digital signature pad, printer, and speaker. These stations allow citizens to initiate certain procedures (e.g. request a police clearance certificate or replace a lost driver's license) without staff assistance, guided by an AI system. The AI can identify the user (via facial recognition and electronic ID), understand the requested service, retrieve necessary personal data, and help complete forms or even print certificates on the spot. The Ministry of Interior's KIOSK Project planned to deploy \href{https://www.parlament.hu/documents/10181/39233854/Infojegyzet_2021_73_mesterseges_intelligencia_a_kozigazgatasban.pdf/#:~:text=A%20KIOSK%20projekt%20keret%C3%A9ben%20k%C3%ADs%C3%A9rleti,darabot%20fog%02nak%20bel%C5%91l%C3%BCk%20orsz%C3%A1gszerte%20a}{\textcolor{blue}{400}} such kiosks nationwide in Government Offices. No information could be found of whether these kiosks where successfully installed. \\
\bottomrule
\end{tabular}
\end{table}

\begin{table}[ht]
\centering
\caption{Assessment of Hungary's AI Strategy: goal 22 (as of February 2025)}
\label{tab:assessment-goals-22}
\footnotesize
\begin{tabular}{>{\raggedright\arraybackslash}p{0.8cm}>{\raggedright\arraybackslash}p{2.5cm}>{\arraybackslash}p{12cm}}
\toprule
\textbf{No.} & \textbf{Goal} & \textbf{Assessment (as of 02.2025)} \\
\midrule
22. & Implementation of smartgrid technologies & The Ministry of Energy announced new \href{https://infostart.hu/gazdasag/2025/01/01/egy-gonddal-kevesebb-az-okosmerosoknek-valtozik-a-rendszer-januar-1-tol}{\textcolor{blue}{regulations}} starting from January 1st of 2025.
\begin{itemize}
\item smart-metered customers no longer need to submit meter readings and can monitor daily consumption online.
\item new time-of-use tariff structure: as of 2025, three intraday periods are defined – daytime (6:00–17:00), peak evening (17:00–22:00), and night off-peak. Initially, the tariff levels for smart-meter users are set equal to standard rates (so no bill shock occurs), but this regulatory change created the conditions for differentiated pricing in the future.
\item consumers with smart meters are being \href{https://infostart.hu/gazdasag/2025/01/01/egy-gonddal-kevesebb-az-okosmerosoknek-valtozik-a-rendszer-januar-1-tol}{\textcolor{blue}{moved}} to a time-series (hourly) settlement system instead of yearly reconciliations.
\end{itemize}

In effect, Hungary now has the technical and regulatory framework to implement dynamic pricing, which can encourage consumers to shift usage to periods of high renewable generation (e.g. midday solar peaks). Concerning solar power, Hungary's capacity has grown– the original 2030 target of 6 GW solar was already met by early 2024, prompting a new \href{https://www.schoenherr.eu/content/hungary-amendments-to-grid-capacity-allocation-rules-may-signal-last-chapter-of-solar-power-gold-rush}{\textcolor{blue}{12 GW solar target}} for 2030.

These efforts were supported by the EU: the \href{https://commission.europa.eu/business-economy-euro/economic-recovery/recovery-and-resilience-facility/recovery-and-resilience-plan-hungary/hungarys-recovery-and-resilience-supported-projects-nation-wide-investment-scheme_en}{\textcolor{blue}{``Classic and Smart Grid Development''}} program under Hungary's RRF plan is the largest of this kind. The HUF 170 billion (\textasciitilde EUR 415,000,000) investment (2022–2026) to be carried out by the transmission system operator (TSO) and all the major DSOs. The goal is to ``allow for the secure and flexible integration'' of up to 3,000 MW of new renewable capacity into the grid. Thanks to such initiatives, private and state-owned companies started to upgrade the existing infrastructure:
\begin{itemize}
\item E.ON Hungária has announced an increase in grid investment – about HUF 160 billion (\textasciitilde EUR 390,000,000) in 2024 alone, and up to HUF 410 billion (\textasciitilde EUR 1,000,000,000) over the coming years – aimed at the \href{https://industrialnews.co.uk/e-on-installing-165000-smart-meters-in-hungary/}{\textcolor{blue}{``complete renewal''}} of the electricity network in its license areas (West and Central Hungary). This includes conventional upgrades (replacing old lines, expanding capacity) and smart technologies.
\item MVM \href{https://www.smart-energy.com/industry-sectors/smart-grid/amp/page/476/?amp-wp-skip-redirect=1}{\textcolor{blue}{secured}} a HUF 50 billion (\textasciitilde EUR 120,000,000) loan from the European Investment Bank to modernize and expand the grid in south-eastern Hungary (the area served by Démász) – this included components of smart metering and grid automation (e.g. substation upgrades and telecontrol).
\end{itemize}

The EU has also major smart grid projects (EU-Funded), with the goal of promoting cross-border smart grid technologies and connectivity.
\begin{itemize}
\item \href{https://danubeingrid.eu/wp-content/uploads/2023/06/202305_Danube_InGrid_leaflet_ENG-1.pdf#:~:text=Promoters%3A%20Z%C3%A1padoslovensk%C3%A1%20distribu%C4%8Dn%C3%A1%2C%20a,ipari%20%C3%81tviteli%20Rendszerir%C3%A1ny%C3%ADt%C3%B3%20Zrt%20Promoters}{\textcolor{blue}{Danube InGrid – First Wave}} (2020–2025): This project focuses on Western Hungary (North Transdanubia) and Western Slovakia. The promoters are E.ON Észak-dunántúli Áramhálózati Zrt. (the DSO in NW Hungary), Slovakia's TSO SEPS, and Západoslovenská distribučná (ZSD, West Slovakia's DSO), with MAVIR supporting. The project's purpose is to strengthen the interaction and integration between the Slovak and Hungarian electricity markets by adopting smart grid technologies at both the national and cross-border interface. Concretely, it involves installing smart devices on high-voltage and medium-voltage lines, constructing or modernizing smart substations, deploying an optical fiber communications network for the grid, and implementing new IT systems for grid management. The first wave (Action 10.7-0008-SKHU) is co-financed by the EU's Connecting Europe Facility (CEF) with a grant of HUF 41 billion (\textasciitilde EUR 102,000,000) and has a total budget of about HUF 120 billion (\textasciitilde EUR 291,000,000).
\item \href{https://danubeingrid.eu/wp-content/uploads/2023/06/202305_Danube_InGrid_leaflet_ENG-1.pdf#:~:text=Promoters%3A%20Z%C3%A1padoslovensk%C3%A1%20distribu%C4%8Dn%C3%A1%2C%20a,ipari%20%C3%81tviteli%20Rendszerir%C3%A1ny%C3%ADt%C3%B3%20Zrt%20Promoters}{\textcolor{blue}{Danube InGrid – Second Wave}} (2022–2029): The second phase extends the concept to Northeastern Hungary and Eastern Slovakia, where renewable integration challenges are also growing. Promoters include MVM ÉMÁSZ (the DSO in NE Hungary) and ELMŰ Hálózati Kft (the DSO for Budapest region, also involved due to network linkages), together with Slovakia's Východoslovenská distribučná (East SK DSO) and SEPS. MAVIR again acts as a supporting partner. This phase focuses on deeper TSO–DSO coordination and advanced data management: designing ``smart substation'' architectures, establishing real-time data exchange platforms between the transmission and distribution level, and implementing joint control schemes for balancing. The budget for wave two is around HUF 100 billion (\textasciitilde EUR 243,000,000) (2022–2029) with CEF expected to co-fund a portion.
\end{itemize} \\
\bottomrule
\end{tabular}
\end{table}

\FloatBarrier

\subsection{Conclusions}
\label{subsec:conclusions}

\subsubsection{Estimated total capital}
\label{subsubsec:estimated-total-capital}

The total known public investment across the AI strategy goals amounts to more than HUF 1,896 billion, equivalent to approximately EUR 4.65 billion (see \emph{Table~\ref{tab:hungary-financial-allocations}}). This amount includes only those goals for which concrete financial data is publicly accessible or referenced in official reports, and includes solely the subset of 22 goals selected from the total of 71.

Based on the assessment of the 22 strategic goals, there is a disparity in budget transparency and investment scale. The available data suggests a strong concentration of funding in a few key infrastructural projects, while numerous goals remain under-documented in terms of financial commitment.

Three goals account for the overwhelming majority of the known financial commitment:

\begin{itemize}
\item Goal 22 – Implementation of Smart Grid Technologies received the largest allocation: HUF 1,051 billion (\textasciitilde EUR 2.56 billion). This investment was primarily supported by Hungary's Recovery and Resilience Facility (RRF) plan, EU infrastructure programs, and national energy operators (e.g., E.ON, MVM).
\item Goal 19 – Development of the Data Wallet and National Digital Identity Platform was allocated HUF 733 billion (\textasciitilde EUR 1.80 billion). This project evolved into a broader e-identity initiative, aligned with the EU's eIDAS 2.0 regulation.
\item Goal 17 – Digitalization of the Healthcare Sector, particularly the national health data infrastructure and related telemedicine services, received HUF 65 billion (\textasciitilde EUR 160 million).
\end{itemize}

These three goals alone represent more than 95\% of the documented financial scope of the strategy.

Some goals were probably merged or implemented under other objectives (Goal 5 and 10), possibly due to strategic or operational overlaps. Despite the extensive scope of the strategy, for 10 out of 22 goals, no publicly available financial information could be found. Their exclusion from financial reporting may suggest that the projects are still in planning phases, their funding comes from indirect sources or private partnerships, or their implementation was postponed or deprioritized. While several goals are linked to official tenders or projects with corresponding reference links, many do not offer publicly traceable information.

\begin{table}[ht]
\centering
\caption{Mapping of government financial allocations to the specific objectives of Hungary's AI Strategy}
\label{tab:hungary-financial-allocations}
\footnotesize
\begin{tabular}{>{\raggedright\arraybackslash}p{1.5cm}>{\raggedright\arraybackslash}p{4cm}>{\raggedright\arraybackslash}p{3cm}>{\arraybackslash}p{3cm}}
\toprule
\textbf{Goal No.} & \textbf{HUF} & \textbf{EUR} & \textbf{Reference link(s)} \\
\midrule
1 & 50,000,000 & 125,000 & \href{https://www.ceginformacio.hu/cr9312108599}{\textcolor{blue}{link}} \\
2 & 9,000,000,000 & 23,000,000 & \href{https://milab.tk.hu/hirek/2022/11/a-projekt-alapadatai}{\textcolor{blue}{link}} \\
3 & 6,000,000,000 & 16,000,000 & \href{https://www.nemzetilaborok.nkfih.gov.hu/digital-transformation/files-to-download-240826/national-laboratory-for-240826}{\textcolor{blue}{link}} \\
4 & 1,000,000,000 & 2,500,000 & \href{https://neum.hu/328-2/}{\textcolor{blue}{link}} \\
5 & \multicolumn{3}{l}{\emph{merged with Goal No. 4}} \\
6 & \multicolumn{3}{l}{\emph{unknown}} \\
7 & 14,000,000,000 & 35,200,000 & \href{https://hirek.unideb.hu/en/komondor-hungarys-most-powerful-supercomputer-awakening}{\textcolor{blue}{link}};  \href{https://kormany.hu/hirek/nyertes-palyazatnak-koszonhetoen-epul-uj-magyar-szuperszamitogep-csillebercen-a-jelenlegi-kapacitas-negyvenszereset-tudja-majd-levente}{\textcolor{blue}{link}} \\
8 & 14,500,000,000 & 40,400,000 & \href{https://admissions.sze.hu/sze-continues-to-expand-its-zalazone-development-base}{\textcolor{blue}{link}};  \href{https://trademagazin.hu/en/elkeszult-magyarorszag-legnagyobb-ontozesfejlesztesi-beruhazasa-mezohegyesen/}{\textcolor{blue}{link}} \\
9 & \multicolumn{3}{l}{\emph{unknown}} \\
10 & \multicolumn{3}{l}{\emph{merged with Goal No. 8}} \\
11 & 1,500,000,000 & 3,900,000 & \href{https://hungarytoday.hu/hungarian-innovation-for-better-analysis-of-medical-findings-using-ai/}{\textcolor{blue}{link}} \\
12 & \multicolumn{3}{l}{\emph{unknown}} \\
13 & \multicolumn{3}{l}{\emph{unknown}} \\
14 & \multicolumn{3}{l}{\emph{unknown}} \\
15 & \multicolumn{3}{l}{\emph{unknown}} \\
16 & \multicolumn{3}{l}{\emph{unknown}} \\
17 & 65,000,000,000 & 160,000,000 & \href{https://okfo.gov.hu/egeszsegugyi-fejlesztesek/hazai-fejlesztesek/2014-2020-fejlesztesi-idoszak/lezarult-fejlesztesek/efop-1-9-6-16-2017-00001}{\textcolor{blue}{link}} \\
18 & \multicolumn{3}{l}{\emph{unknown}} \\
19 & 733,000,000,000 & 1,802,500,000 & \href{https://neum.hu/en/328-2/}{\textcolor{blue}{link}}; \href{https://www.palyazat.gov.hu/hirek/dimop-plusz-uj-lendulet-magyarorszag-digitalis-fejlodesehez}{\textcolor{blue}{link}} \\
20 & 1,000,000,000 & 2,500,000 & \href{https://www.palyazat.gov.hu/programok/szechenyi-2020/efop/efop-3213-17/alapadatok}{\textcolor{blue}{link}} \\
21 & \multicolumn{3}{l}{\emph{unknown}} \\
22 & 1,051,000,000,000 & 2,561,000,000 & \href{https://commission.europa.eu/business-economy-euro/economic-recovery/recovery-and-resilience-facility/recovery-and-resilience-plan-hungary/hungarys-recovery-and-resilience-supported-projects-nation-wide-investment-scheme_en}{\textcolor{blue}{link}}; \href{https://industrialnews.co.uk/e-on-installing-165000-smart-meters-in-hungary/}{\textcolor{blue}{link}}; \href{https://www.smart-energy.com/industry-sectors/smart-grid/amp/page/476/?amp-wp-skip-redirect=1}{\textcolor{blue}{link}}; \href{https://danubeingrid.eu/wp-content/uploads/2023/06/202305_Danube_InGrid_leaflet_ENG-1.pdf}{\textcolor{blue}{link}} \\
\midrule
\textbf{Total} & \textbf{1,896,050,000,000} & \textbf{4,647,125,000} & \\
\bottomrule
\end{tabular}
\end{table}

\subsubsection{Time keeping}
\label{subsubsec:time-keeping}

The National AI Strategy presented a structured timeline, targeting the 2020–2030 period. While several foundational programs were initiated as scheduled, the implementation pace of many others proved inconsistent or difficult to verify. A notable trend throughout the assessment is that numerous goals were pursued as several standalone initiatives or have instead merged into pre-existing or parallel projects. In other cases, such as Goal 13 (AI research cluster in public administration), no independent project was launched; rather, existing institutions like the AI Coalition and National AI Laboratory were informally tasked with fulfilling the objective.

Despite the shortcomings in timekeeping and transparency, the assessment also reveals that a considerable number of goals have been completed, and some were even implemented ahead of schedule. For instance, Goal 4 (AI Accelerator Centre) had its funding published in January 2021—six months earlier than planned—and physical centers were established in Debrecen and Zalaegerszeg. Similarly, Goal 7 (Enhancing supercomputer capacities) resulted in the operationalization of the komodor supercomputer by late 2022, meeting its performance target of 5 petaflops. Although government announcements are often lacking, several functioning services indicate that implementation occurred in practice. For example, the MIA chatbot and self-service kiosks linked to Goal 21 are already in use across Hungary's public administration system, handling millions of queries. Likewise, Goal 19's data wallet initiative has matured into a national digital identity platform, with integration capabilities available to third-party developers as of early 2025. These outcomes demonstrate that despite limited visibility and incomplete reporting, multiple elements of the strategy have been translated into operational solutions.

Despite the strategy mandating biannual reviews, no formal update has been published as of February 2025. This lack of structured follow-up has contributed to a fragmented implementation landscape, where responsibilities and project timelines can blur across initiatives. Although initial planning was comprehensive, the absence of iterative, time-adjusted action plans (the majority of the goals had to be completed by 2022/2023, with no further actions) and the limited transparency around ongoing progress signal a need for stronger coordination, updated targets, and continuous monitoring to ensure the strategy's long-term relevance and impact. These factors contribute to the impression that the original strategy may have been pursued primarily to comply with European Union expectations, rather than out of sustained national commitment. However, in a potentially significant shift, the Hungarian government appointed the former Minister for Innovation and Technology as Government Commissioner for Artificial Intelligence in February 2025. This recent move signals a renewed political commitment to AI policy and may lay the groundwork for a revised or second-phase strategy that addresses the gaps of the first.

\subsubsection{Insights from Tibor Gulyás}
\label{subsubsec:insights-gulyas}

The author had the opportunity to conduct an interview with Tibor Gulyás, who currently serves as Chief Strategic Advisor to Dr. László Palkovics, Government Commissioner for AI. From 2018 to 2022, Mr. Gulyás held the position of Deputy State Secretary for Innovation at the Ministry for Innovation and Technology (ITM), where he worked closely with Dr. Palkovics and played a key role in shaping Hungary's innovation policy, including the development of the national AI strategy. Following the transformation of the ITM into the Ministry of Technology and Industry, and after Dr. Palkovics's resignation and the creation of the Energy Ministry, Gulyás continued to serve as Deputy State Secretary for Strategic Affairs in both ministries. In February 2025, he was appointed to his current advisory role.

According to him, the origins of Hungary's national AI strategy can be traced back to 2018, when AI emerged as one of several rapidly rising technology topics within the broader governmental innovation agenda. He recalled that emerging fields such as AI, hydrogen technologies, and others were often first treated as thematic novelties, and the ministry sought to institutionalize strategic responses through what Gulyás described as ``platformization''—a structuring process that brought together academic, governmental, and corporate actors into coordinated networks. It was within this framework that the AI Coalition was created, serving as a hub for gathering input and facilitating cross-sector collaboration. Gulyás emphasized that the Hungarian strategy was not a local adaptation of EU narratives but originated from domestic recognition of AI's long-term economic and technological significance. He described that early efforts to respond to the opportunity included the initiation of a significant EU-funded call to support AI-related programs. The process of strategy formulation itself involved a wide range of contributors and was shaped by a combination of proactive government vision and responsive engagement with emerging stakeholder needs from industry and academia. However, when it came to the specific goals, Gulyás also acknowledged certain limitations in the strategy's composition. The drafting process brought together a broad range of contributors and interest groups, which, while ensuring wide representation, also led to the inclusion of some narrower stakeholder agendas alongside the more unified, consensus-driven objectives. Furthermore, Gulyás highlighted that Hungary was one of the earliest countries in the European Union to publish a national AI strategy, underscoring its early commitment to the field.

Despite the strategy's early momentum, Tibor Gulyás noted that external crises caused a major challenge as they disrupted governmental priorities and diverted resources. He considered the COVID-19 pandemic to have had a paradoxical but mostly advantageous effect: while some AI-related funding streams were reallocated to urgent health and economic responses, the pandemic also accelerated digitalization processes across the public and private sectors, creating a better ground for AI adoption. In contrast, the energy crisis and the outbreak of war in Ukraine in 2022 had a negative impact. These events, according to Gulyás, reshaped the state's agenda, pushed AI down the list of priorities, as ensuring energy supply became a dominant concern. This shift led to the creation of the Ministry of Energy and re-channeled administrative capacity and financial resources toward energy-related crisis management.

Gulyás also emphasized that implementation benefited from the institutional setup in place between 2020 and the end of 2022. During this period, the Ministry for Innovation and Technology (ITM), under the leadership of Minister Dr. László Palkovics, served as the central authority for both the formulation and execution of the strategy. He highlighted that the strategic advantage of this arrangement lay in the structure of decision-making power, resource allocation, and sectoral oversight within a single ministry. Gulyás explicitly rejected the notion that concentrating responsibility within one ministry would have posed disadvantages, arguing instead that it strengthened implementation by assigning the mandate to a ministry that already possessed the relevant competencies. This concentration, in his opinion, enabled coordination, accountability, and efficient deployment of funding. He also noted the important roles played by key figures within the ministry, including Solymár Károly, the Deputy State Secretary for Infocommunications, who maintained close ties with the AI Coalition from the strategy's formulation through its implementation.

Following the resignation of Minister Palkovics in late 2022, the previously centralized structure was restructured as part of a broader government reorganization. Gulyás explained that the Ministry of Technology and Industry ceased to exist, and responsibilities tied to the AI strategy were redistributed across multiple ministries and institutions. Portions of the strategy were absorbed by the newly created Ministry of Energy, where Solymár Károly continued his work as Deputy State Secretary. Oversight of the Digital Renewal Program (DIMOP) was linked to the Government Office, which also became involved in AI-related implementation. In early 2023, with the formation of the Ministry for National Economy, Szabolcs Szolnoki was appointed Deputy State Secretary for Technology. He placed strong emphasis on implementing the strategy and emerged as a key figure in its execution. Gulyás views the appointment of a dedicated government commissioner for AI in 2025 as a recognition by the Prime Minister of the field's rising importance. In his view, the decision was driven by the realization that AI had become a matter of such strategic relevance that it required an official directly responsible for this field, assigned to the PMO, with the authority to move things forward. In Gulyás' opinion, the fragmentation of the AI-related duties from 2022 did not cause the implementation to halt or to weaken.

Gulyás explained that the announcement of the new AI strategy is only a matter of days. He noted that the previous strategy had been thoroughly reviewed, its effects were measured, and its goals reassessed. He rejected the claim that the AI Coalition had ceased to exist and emphasized that it continues to play a significant role in shaping the forthcoming strategy. Gulyás emphasized that the upcoming version of the strategy is not a break from the previous one, but a refinement based on experience. Therefore, the pyramid framework (foundation pillars; sector specific focus areas; and transformative programs) will remain, and most changes will focus on the transformative programs. It will place greater emphasis on demonstrating that AI can bring benefits to both society and the business sector. New institutional leadership will now rely on a two-tier system: the government commissioner for AI will oversee overall strategic alignment and interministerial coordination, while there will be designated AI officers within each ministry, who will be responsible for sector-specific execution.

\subsubsection{Insights from Dr. Roland Jakab}
\label{subsubsec:insights-jakab}

The author had also the opportunity to meet and talk with Dr. Roland Jakab, President of the AI Coalition of Hungary, during a closed, small-group discussion hosted by TechStage \cite{techstage}, a student-led initiative aimed at connecting university students with leading figures in the domestic technology sector. Dr. Jakab has been a key figure (alongside former Minister of Innovation and Technology Dr. László Palkovics) in the development of the AI Strategy.

Dr. Jakab discussed that although the National AI Strategy was developed with a strong conceptual foundation by the Coalition, its implementation has faced significant challenges. In his view, the lack of institutional accountability and inter-ministerial coordination has hindered the strategy's execution. He emphasized the absence of a dedicated body with both a clear mandate and the necessary authority to oversee progress and ensure alignment across governmental entities. As a result, the application of the strategy has been fragmented.

Dr. Jakab also noted that the strategy predates the widespread adoption of large language models (LLMs), such as ChatGPT, and therefore does not adequately reflect the current technological landscape. He suggested that the strategy requires refinement to incorporate these transformative developments in AI.

Despite the shortcomings, Dr. Jakab highlighted the Data Wallet initiative as a significant achievement. He pointed out that Hungary was the first country in the region to implement such a solution, which was subsequently integrated into the broader Digital Public Administration program (Digitális Állampolgárság Program), reflecting a promising example of digital innovation within the region.
\newpage
\section{Singapore's national AI strategy}
\label{sec:singapore-ai-strategy}

\subsection{Timeframe and responsible governmental body}
\label{subsec:singapore-timeframe-responsible-body}

Singapore's national efforts to foster artificial intelligence are structured around two publicly accessible strategic documents:

\begin{itemize}
\item National AI Strategy 1.0 (NAIS 1.0), published in November 2019 \cite{nais2019}, and
\item National AI Strategy 2.0 (NAIS 2.0), published in December 2023 \cite{nais2023}.
\end{itemize}

Both strategies were formulated and are overseen by the Smart Nation and Digital Government Office (SNDGO), which operates under the Prime Minister's Office (PMO). The SNDGO works in close collaboration with the Ministry of Communications and Information (MCI) and relevant implementation agencies, most notably AI Singapore (AISG) and the Infocomm Media Development Authority (IMDA). Shortly before NAIS 2.0, the SNDGG merged with the digital development functions of the Ministry of Communications and Information (MCI) to form an enlarged Smart Nation Group.

Both strategies emphasize cross-agency collaboration and position the SNDGO as the central coordinating body. The National AI Office (later renamed National AI Group), nested within SNDGO, oversees strategic alignment, while AI Singapore, hosted at the National University of Singapore (NUS), remains the key implementation agency. Other ministries and statutory boards (e.g., Ministry of Health, Ministry of Education, HTX – Home Team Science and Technology Agency) execute sector-specific programs under their jurisdiction.

\subsection{Targeted sectors / areas to be supported}
\label{subsec:singapore-targeted-sectors}

Both NAIS 1.0 and NAIS 2.0 identify targeted areas for AI adoption and development. NAIS 1.0 outlines project and enablers, the second designed to strengthen the ecosystem to drive AI innovation and adoption:

\textbf{NAIS 1.0 (2019–2023) targeted sectors:}
\begin{itemize}
\item Transport and Logistics
\item Smart Cities and Estates
\item Healthcare
\item Education
\item Border Security
\end{itemize}

These five sectors were supported by five enabling pillars: data architecture, research and talent development, governance frameworks, multi-stakeholder partnerships, and international collaboration.

NAIS 2.0 expands significantly upon the earlier version, incorporating broader societal, scientific, and governance dimensions. It also emphasizes thinking in domains instead of specific projects, and also switching from local to international ambitions.

\textbf{NAIS 2.0 (2023–2028) domains:}
\begin{itemize}
\item Activity Drivers (Research, Industry and Government)
\item People and Communities (Talent, Capabilities and Placemaking)
\item Infrastructure and Environment (Compute, Data, Trusted Environment and Leadership in Thought and Action)
\end{itemize}

\subsection{Goals and actions}
\label{subsec:singapore-goals-actions}

\subsubsection{NAIS 1.0}
\label{subsubsec:nais-1-0}

\paragraph{Projects}
\label{para:nais-1-0-projects}

See \emph{Table~\ref{tab:nais-1-0-projects}}.

\begin{table}[ht]
\centering
\caption{NAIS 1.0 Projects in Singapore's AI Strategy}
\label{tab:nais-1-0-projects}
\footnotesize
\begin{tabular}{>{\raggedright\arraybackslash}p{1cm}>{\centering\arraybackslash}p{3cm}>{\arraybackslash}p{8cm}>{\arraybackslash}p{1.5cm}}
\toprule
\textbf{No.} & \textbf{Goal} & \textbf{Description} & \textbf{Deadline} \\
\midrule
23 & \multirow{3}{3cm}{\centering Intelligent Freight Planning\\\textit{(Category: Transport and Logistics)}} & AI-supported planning: Develop a common and trusted data platform for integrated freight and truck movement across ports, roads, and depots. & 2022 \\
\cmidrule{1-1}\cmidrule{3-4}
24 & & Dynamic assignment: Use AI to pool and assign freight jobs dynamically, improving asset utilization and job efficiency. & 2025 \\
\cmidrule{1-1}\cmidrule{3-4}
25 & & Urban logistics optimization: Enable government agencies to use AI models for better urban and infrastructure planning. & 2030 \\
\midrule
26 & \multirow{3}{3cm}{\centering Seamless and Efficient Municipal Services\\\textit{(Category: Smart Cities and Estates)}} & AI-powered chatbot: Implement AI chatbots to guide citizens during municipal issue reporting and route feedback automatically to the right agency. & 2022 \\
\cmidrule{1-1}\cmidrule{3-4}
27 & & Sensor-driven maintenance: Deploy AI to process sensor data and predict faults in lifts and infrastructure for preventive action. & 2025 \\
\cmidrule{1-1}\cmidrule{3-4}
28 & & Resident-driven planning: Analyze public usage data for better planning of shared facilities, tailored to community needs. & 2030 \\
\midrule
29 & \multirow{3}{3cm}{\centering Chronic Disease Prediction and Management\\\textit{(Category: Healthcare)}} & AI diagnostics: Deploy AI like SELENA+ to analyze medical images (e.g., retina scans) and predict risks of chronic conditions with high accuracy. & 2022 \\
\cmidrule{1-1}\cmidrule{3-4}
30 & & Care team support: Use AI to help primary care teams prioritize patient interventions based on risk profiles and progress tracking. & 2025 \\
\cmidrule{1-1}\cmidrule{3-4}
31 & & Self-management tools: Provide AI tools for patients to monitor their conditions and receive personalized alerts and lifestyle nudges. & 2030 \\
\midrule
32 & \multirow{3}{3cm}{\centering Personalized Education through Adaptive Learning\\\textit{(Category: Education)}} & Adaptive learning: Integrate machine learning into national education platforms (e.g., Student Learning Space) to tailor lesson content to students' individual progress. & 2022 \\
\cmidrule{1-1}\cmidrule{3-4}
33 & & Automated marking: Use AI-based tools to assess open-ended answers, freeing up teacher time and enabling faster feedback. & 2025 \\
\cmidrule{1-1}\cmidrule{3-4}
34 & & Learning companions: Develop AI-driven companions to engage and motivate students, recommend activities, and encourage self-reflection. & 2030 \\
\midrule
35 & \multirow{3}{3cm}{\centering Border Clearance Operations\\\textit{(Category: Border Security)}} & Pre-arrival risk profiling: Use AI to assess passenger data in advance (e.g., arrival cards, airline data) and tier security checks accordingly. & 2022 \\
\cmidrule{1-1}\cmidrule{3-4}
36 & & Seamless clearance: Enable 100\% automated, biometric-based clearance including for first-time travelers, reducing queue times. & 2025 \\
\cmidrule{1-1}\cmidrule{3-4}
37 & & Security enhancement: Apply AI to continuously monitor clearance processes and flag anomalies for investigation, improving border safety. & 2030 \\
\bottomrule
\end{tabular}
\end{table}

\paragraph{Enablers}
\label{para:nais-1-0-enablers}

See \emph{Table~\ref{tab:nais-1-0-enablers}} and \emph{Table~\ref{tab:nais-1-0-enablers-2.0}}.
\begin{table}[ht]
\centering
\caption{NAIS 1.0 Enablers in Singapore's AI Strategy: no. 38 - 46}
\label{tab:nais-1-0-enablers}
\footnotesize
\begin{tabular}{>{\raggedright\arraybackslash}p{1cm}>{\centering\arraybackslash}p{3cm}>{\arraybackslash}p{8cm}>{\arraybackslash}p{2.5cm}}
\toprule
\textbf{No.} & \textbf{Enabler} & \textbf{Description} & \textbf{Deadline} \\
\midrule
38 & \multirow{4}{3cm}{\centering Triple Helix Partnership} & Strengthen R\&D collaboration between research institutions, industry, and government. SGD 200,000,000 (\textasciitilde EUR 138,000,000) investment to enhance supercomputing and network infrastructure (from 1 to 15–20 petaflops), to boosts national research and AI capability. & 2024 \\
\cmidrule{1-1}\cmidrule{3-4}
39 & & Promote AI uptake, especially in SMEs, through toolkits, training programs, and AI Centres of Excellence. Support 100+ companies in adopting AI and engage thousands of workers in AI capability development. & 2023 \\
\cmidrule{1-1}\cmidrule{3-4}
40 & & Strengthen collaboration between research institutions and industry through over 15 existing AI public–private partnerships. Expand initiatives like joint labs, product co-development, and shared training. Flagship initiatives include the Singtel–NTU AI Lab and IMDA–SUTD Smart Estates Partnership, targeting smart infrastructure, healthcare, and manufacturing. & Not mentioned \\
\cmidrule{1-1}\cmidrule{3-4}
41 & & Launch regulatory sandboxes and testbeds (e.g., Punggol Digital District) to develop explainable, safe AI in real-world settings. Linked to Model AI Governance Framework v2.0, with sector-specific implementations and global benchmarking efforts. & Not mentioned \\
\midrule
42 & \multirow{3}{3cm}{\centering AI Talent and Education} & Expand the AI Apprenticeship Program (AIAP) to train 500 AI engineers. Launch postgraduate AI scholarships in partnership with industry (e.g. Grab, Nvidia, Alibaba). Support career conversion and upskilling via TeSA (TechSkills Accelerator). & 2025 \\
\cmidrule{1-1}\cmidrule{3-4}
43 & & Scale AI literacy programs: AI for Everyone, AI for Students, AI for Kids. Train 100,000 citizens and 25,000 professionals in basic AI implementation. & 2025 \\
\cmidrule{1-1}\cmidrule{3-4}
44 & & Launch Tech@SG to ease hiring of top AI experts. Partner with IBM to train 2,500 Singaporeans through formal AI curricula. Focus on building local capabilities via partnerships with global leaders. & 2025 \\
\midrule
45 & \multirow{2}{3cm}{\centering Data Architecture} & Implement a Trusted Data Sharing Framework (by IMDA) to guide companies on legal, regulatory, and technical safeguards. Launch a Public-Private Data Sharing Framework to define conditions for government data sharing with non-government entities. These frameworks support National AI Projects by enabling responsible and secure cross-sectoral data access. & Not mentioned \\
\cmidrule{1-1}\cmidrule{3-4}
46 & & Identify sector-specific trusted data intermediaries to facilitate data fusion, anonymization, and secure sharing. These can be public or private entities depending on the context. The initiative supports AI R\&D and innovation by creating reusable data access pipelines. & Not mentioned \\
\bottomrule
\end{tabular}
\end{table}

\begin{table}[ht]
\centering
\caption{NAIS 1.0 Enablers in Singapore's AI Strategy: no. 47 - 50}
\label{tab:nais-1-0-enablers-2.0}
\footnotesize
\begin{tabular}{>{\raggedright\arraybackslash}p{1cm}>{\centering\arraybackslash}p{3cm}>{\arraybackslash}p{8cm}>{\arraybackslash}p{2.5cm}}
\toprule
\textbf{No.} & \textbf{Enabler} & \textbf{Description} & \textbf{Deadline} \\
\midrule
47 & \multirow{2}{3cm}{\centering Progressive and Trusted Environment} & Implement the Model AI Governance Framework (v1 published in 2019). Develop sector-specific AI governance codes, technical solutions for explainable AI, and certification programs in AI ethics. Engage universities and research centres (e.g. SMU, SUTD, Live with AI) to study AI's long-term impact on society, jobs, and institutions. & Not mentioned \\
\cmidrule{1-1}\cmidrule{3-4}
48 & & Launch IPOS (Intellectual Property Office of Singapore) International to provide tailored intellectual property strategies for AI companies. Review Singapore's IP legislation to support commercialization of AI innovations. & Not mentioned \\
\midrule
49 & \multirow{2}{3cm}{\centering International Collaboration} & Enhance collaboration with WEF, OECD, ISO, IEC to shape global norms on AI. Work with the WEF Centre for the Fourth Industrial Revolution to enhance and promote adoption of the Model AI Governance Framework, incorporating global industry feedback and use cases. & Not mentioned \\
\cmidrule{1-1}\cmidrule{3-4}
50 & & Lead multi-site AI testbeds in partnership with international research institutes. A notable, already example is the 2018 MoU with French institutes (INRIA, CNRS, INSERM) covering 5 research areas including federated learning, explainable AI, and AI in healthcare. & Not mentioned \\
\bottomrule
\end{tabular}
\end{table}
\subsubsection{NAIS 2.0}
\label{subsubsec:nais-2-0}

In contrast to the original NAIS 1.0 strategy, where thematic focus areas were outlined more broadly, the updated NAIS 2.0 introduces a significantly more structured framework by directly linking each enabler to at least one concrete action. This alignment ensures that strategic intent is operationalized through measurable initiatives. By explicitly connecting enablers to targeted actions, the new approach is designed with the goal to generate compounded added value across the ecosystem. There are 14 actions linked to 10 enablers, as demonstrated at \emph{Table~\ref{tab:nais-2-0-enablers-actions}}, \emph{Table~\ref{tab:nais-2-0-enablers-actions-2}} and \emph{Table~\ref{tab:nais-2-0-enablers-actions-3}},.

\begin{table}[ht]
\centering
\caption{NAIS 2.0 Enablers and Actions in Singapore's AI Strategy: no. 57 - 62}
\label{tab:nais-2-0-enablers-actions}
\footnotesize
\begin{tabular}{>{\raggedright\arraybackslash}p{0.5cm}>{\centering\arraybackslash}p{2cm}>{\arraybackslash}p{3cm}>{\arraybackslash}p{3cm}>{\arraybackslash}p{5.5cm}}
\toprule
\textbf{No.} & \textbf{Enabler} & \textbf{Description} & \textbf{Associated action} & \textbf{Action description} \\
\midrule
51 & \multirow{2}{2cm}{\centering Industry} & \multirow{2}{3cm}{Adopt a sector-specific, use case-centric approach, and focus initial efforts on leading economic sectors assessed to be most ready for AI-driven transformation.} & Anchor new AI Centres of Excellence (CoEs) across companies, and explore establishing Sectoral AI CoEs to drive sophisticated AI value creation and usage in key sectors & Attract and anchor new AI CoEs in Singapore-based companies that are leading-edge producers (e.g. American Express Decision Science CoE) and sophisticated end-users, in order to conduct value creation activities across the AI stack. Establish a new model of Sectoral AI CoEs to intensify sophisticated AI value creation and usage in selected economic sectors. \\
\cmidrule{1-1}\cmidrule{4-5}
52 &  &  & Strengthen the AI start-up ecosystem, including attracting AI-focused accelerator programs to spur rapid AI experimentation & Strengthen the AI start-up ecosystem by attracting venture builders and AI-focused accelerator programs. These programs, led by tech companies and VCs, will provide capital, technical and business expertise, infrastructure, and networks. The aim is to nurture globally oriented, AI-native start-ups and accelerate AI value discovery and experimentation across industries. \\
\midrule
53 & Government & Coordinate public sector innovation, integrate AI into functional domains, and demonstrating sectoral leadership. & Improve public service productivity, with new value propositions for citizens & AI adoption will be scaled across domain-specific sectors (e.g. Healthcare, Education) and whole-of-government functions (e.g. HR, Finance, Service Delivery). Central resources such as funding, cross-agency data sharing, and compute infrastructure will support implementation. Public service AI capabilities will be strengthened through structured training programs across workforce segments. \\
\midrule
54 & Research & Foster research institutions, as they underpin Singapore's long-term AI competitiveness, contributing to global knowledge production and local innovation. & Update national AI R\&D plans to sustain leadership in select research areas & Update national AI R\&D plans along five dimensions: research priorities, industry-academia collaboration, recruitment of top researchers, compute infrastructure, and international partnerships. Offer up to SGD 330,000 (\textasciitilde EUR 226,000) per project – in total SGD 33,000,000 (\textasciitilde EUR 22,600,000) through the AISG 100 Experiments (100E) program to co-fund industry-academic AI collaborations. Recent and example initiatives include a joint AI research grant call with Korea in 2023 and the Singapore Conference on AI (SCAI) to convene leading global researchers \\
\midrule
55 & \multirow{2}{2cm}{\centering Talent} & \multirow{2}{3cm}{Develop and attract AI talent. Focus areas include world-class AI researchers (creators) and applied practitioners to implement AI solutions at scale.} & Attract world-class AI creators to work from and with Singapore. & A dedicated team will be established to identify, engage, and support globally recognized AI experts from both academia and industry. The aim is to embed top AI talent into Singapore's innovation system through bespoke arrangements, including hybrid roles, part-time appointments, and institutional partnerships. These creators are expected to catalyze local research and product development \\
\cmidrule{1-1}\cmidrule{4-5}
56 &  &  & Boost AI practitioner pool to 15,000 & Singapore aims to grow the national AI practitioner base to 15,000 individuals through redesigned training such as the AI Apprenticeship Programme (AIAP), industry-led attachments, and continuing education programs. \\
\bottomrule
\end{tabular}
\end{table}

\begin{table}[ht]
\centering
\caption{NAIS 2.0 Enablers and Actions in Singapore's AI Strategy: no. 63 - 64}
\label{tab:nais-2-0-enablers-actions-2}
\footnotesize
\begin{tabular}{>{\raggedright\arraybackslash}p{0.5cm}>{\centering\arraybackslash}p{2cm}>{\arraybackslash}p{3cm}>{\arraybackslash}p{3cm}>{\arraybackslash}p{5.5cm}}
\toprule
\textbf{No.} & \textbf{Enabler} & \textbf{Description} & \textbf{Associated action} & \textbf{Action description} \\
\midrule
57 & \multirow{2}{2cm}{\centering Capabilities} & \multirow{2}{3cm}{Build broad-based AI capabilities to transform enterprises and safeguard workforce relevance. Support mechanisms include digital readiness tools, enterprise solution matching, and sector-specific training.} & Intensify enterprise AI adoption for industry transformation & AI adoption will be scaled across enterprises using tools like the AI Readiness Index (AIRI) and support schemes such as SMEs Go Digital and CTO-as-a-Service (CTOaaS). Enterprises will be guided from foundational readiness to tailored, AI-enabled transformation projects, with advanced firms supported through the Digital Leaders Programme (DLP) \\
\cmidrule{1-1}\cmidrule{4-5}
58 &  &  & Upskill workforce through sector-specific AI training programs & Sector leads under the Industry Transformation Maps (ITMs) will develop AI training aligned with Jobs Transformation Maps (JTMs). These will identify how AI affects job roles and guide the design of targeted reskilling programs to prepare workers across 23 industry clusters \\
\midrule
59 & Placemaking & Build a collaborative, visible, and vibrant AI ecosystem by physically and digitally connecting creators, practitioners, and users. & Establish an iconic AI site to co-locate AI creators and practitioners, and nurture the AI community & A dedicated hub will bring together entrepreneurs, researchers, and engineers in shared spaces designed for collaboration. The site will host events, hackathons, and community activities, and will be supported by a digital platform to extend participation beyond physical boundaries \\
\midrule
60 & Compute & Scale access to high-performance computing to enable AI experimentation, innovation, and deployment across industry, research, and the public sector. & Significantly increase high-performance computing available in Singapore & Local computing capacity will be expanded by deepening partnerships with major compute providers, securing GPU-equipped infrastructure, and allocating sufficient carbon and energy budgets. A portion of these resources will be reserved for public good and capability-building use cases. The long-term goal is to transition toward sustainable, net-zero data centers powered by renewables. \\
\midrule
61 & \multirow{2}{2cm}{\centering Data} & \multirow{2}{3cm}{Enhance the availability, quality, and safe use of data for AI development, including privacy-preserving and cross-sector data sharing.} & Build capabilities in data services and privacy-enhancing technologies & Capabilities will be developed in Privacy-Enhancing Technologies (PETs), including synthetic data, federated learning, and homomorphic encryption. The PET Sandbox, launched in 2022, provides funding and regulatory support for companies to pilot privacy-preserving data projects in real-world settings \\
\cmidrule{1-1}\cmidrule{4-5}
62 &  &  & Unlock Government data for use cases that serve the public good & A "data concierge" will be introduced to facilitate access to public sector datasets, broker data-sharing agreements, and support AI developers with data discovery. For selected use cases, the government may also mediate private sector data sharing. These efforts are guided by a Public Good–oriented evaluation framework \\
\bottomrule
\end{tabular}
\end{table}

\begin{table}[ht]
\centering
\caption{NAIS 2.0 Enablers and Actions in Singapore's AI Strategy: no. 51 - 56}
\label{tab:nais-2-0-enablers-actions-3}
\footnotesize
\begin{tabular}{>{\raggedright\arraybackslash}p{0.5cm}>{\centering\arraybackslash}p{2cm}>{\arraybackslash}p{3cm}>{\arraybackslash}p{3cm}>{\arraybackslash}p{5.5cm}}
\toprule
\textbf{No.} & \textbf{Enabler} & \textbf{Description} & \textbf{Associated action} & \textbf{Action description} \\
\midrule
63 & \multirow{2}{2cm}{\centering Trusted Environment} & \multirow{2}{3cm}{Create a trusted environment for AI by embedding governance, assurance, and safety into AI development, deployment, and usage.} & Ensure a fit-for-purpose regulatory environment for AI & Regulatory frameworks will be updated to address novel risks, including those posed by Generative AI. Key instruments such as the Model AI Governance Framework and AI Verify will be reviewed and refined to provide baseline guidance. New initiatives include a common platform for regulators, sector-specific sandboxes, and the expansion of the AI Verify Foundation, which currently includes over 90 corporate members globally. \\
\cmidrule{1-1}\cmidrule{4-5}
& & & Raise security and resilience baseline for AI & The security and resilience of AI systems will be enhanced through updated cybersecurity toolkits, red teaming initiatives, and joint technical guideline development with the private sector. These efforts aim to mitigate misuse such as scams and disinformation, and foster a Testing, Inspection, and Certification (TIC) ecosystem for AI safety assurance \\
\midrule
64 & Leader in Thought and Action & Strengthen Singapore's global credibility and influence by contributing to international AI standards, dialogue, and partnerships. & Establish Singapore as an ambitious and pragmatic international partner on AI innovation and governance & Efforts will focus on deepening bilateral and multilateral relationships, increasing global awareness of Singapore's practical AI governance tools (e.g. AI Verify), and contributing to inclusive, rules-based AI norms. Singapore will continue co-organizing major convenings (e.g. ATxAI, SCAI), chairing the Forum of Small States (FOSS), and developing capacity-building initiatives for its 108 member countries \\
\bottomrule
\end{tabular}
\end{table}

\FloatBarrier
\subsection{Conclusions}
\label{subsec:singapore-conclusions}

\subsubsection{Estimated total capital}
\label{subsubsec:singapore-estimated-total-capital}

Both NAIS 1.0 and NAIS 2.0 are financially supported under Singapore's Research, Innovation and Enterprise (RIE) Fund, the nation's principal mechanism for strategic science and technology investment. The RIE2020 cycle covered the period 2016–2020, with an allocated budget of SGD 19 billion (\textasciitilde EUR 13 billion), while the RIE2025 cycle runs from 2021 to 2025, with a significantly increased budget of SGD 25 billion (\textasciitilde EUR 17 billion), representing approximately 1\% of Singapore's GDP.

The strategies themselves do not itemize specific financial allocations per initiative, and official government statements and public investment figures offer only limited insight into the scale of funding directed toward AI-related programs. These include not only core R\&D efforts under AI Singapore (AISG) but also broader investments in compute infrastructure, talent development, and governance frameworks.

The total amount of directly identifiable investments linked to the actions under NAIS 1.0 and NAIS 2.0 is approximately SGD 2.587 billion (\textasciitilde EUR 1.78 billion), as detailed in the corresponding funding table (see \emph{Table~\ref{tab:singapore-financial-allocations}}).

Among these, the largest single investments include:

\begin{itemize}
\item SGD 1 billion (\textasciitilde EUR 690,000,000) in additional AI funding announced in conjunction with NAIS 2.0, covering compute infrastructure, talent, and ecosystem support.
\item SGD 880 million (\textasciitilde EUR 607,000,000) invested across multi-action clusters including AI research and sectoral integration.
\item SGD 500 million (\textasciitilde EUR 345,000,000) dedicated to high-performance compute resources for AI.
\end{itemize}

Despite these significant sums, it is important to note that many actions under NAIS 2.0 are not explicitly linked to a disclosed funding amount, either due to their integrative nature across ministries or because they are embedded within broader digital or public service programs. The RIE2020 and RIE2025 collectively represents a national R\&D investment of SGD 44 billion equivalent of around EUR 30.3 billion from 2016 to 2025. While these sums cover all domains of science and innovation, a considerable portion is attributed to AI-related activities, especially within the Smart Nation and Digital Economy (SNDE) domain.

\begin{table}[ht]
\centering
\caption{Mapping of government financial allocations to the specific objectives of Singapore's NAIS 1.0 and 2.0}
\label{tab:singapore-financial-allocations}
\footnotesize
\begin{tabular}{>{\raggedright\arraybackslash}p{1.5cm}>{\raggedright\arraybackslash}p{4cm}>{\raggedright\arraybackslash}p{3cm}>{\arraybackslash}p{2.5cm}}
\toprule
\textbf{Goal No.} & \textbf{SGD} & \textbf{EUR} & \textbf{Reference link(s)} \\
\midrule
23-37 & \multicolumn{3}{l}{\emph{unknown}} \\
38 & 880,000,000 & 607,000,000 & \href{https://www.brookings.edu/wp-content/uploads/2021/10/Strengthening-International-Cooperation-AI_Oct21.pdf#:~:text=million%20for%20AI%20research%20through,goals%2C%20AISG%20has%20encouraged%20public}{\textcolor{blue}{link}} \\
39-43 & \multicolumn{3}{l}{\emph{unknown}} \\
44 & 180,000,000 & 124,000,000 & \href{https://www.smartnation.gov.sg/media-hub/press-releases/new-ai-programmes-2021/}{\textcolor{blue}{link}} \\
45-54 & \multicolumn{3}{l}{\emph{unknown}} \\
55 & 7,000,000 & 4,800,000 & \href{https://www.mddi.gov.sg/media-centre/press-releases/ai-initiatives-launched-to-uplift-sg-economic-potential/}{\textcolor{blue}{link}} \\
56 & 20,000,000 & 13,800,000 & \href{https://www.mddi.gov.sg/media-centre/press-releases/ai-initiatives-launched-to-uplift-sg-economic-potential/}{\textcolor{blue}{link}} \\
57-59 & \multicolumn{3}{l}{\emph{unknown}} \\
60 & 500,000,000 & 345,000,000 & \href{https://www.mddi.gov.sg/media-centre/press-releases/ai-initiatives-launched-to-uplift-sg-economic-potential/}{\textcolor{blue}{link}} \\
61-64 & \multicolumn{3}{l}{\emph{unknown}} \\
allocated in addition & 1,000,000,000 & 690,000,000 & \href{https://www.mddi.gov.sg/media-centre/press-releases/ai-initiatives-launched-to-uplift-sg-economic-potential/}{\textcolor{blue}{link}} \\
\midrule
\textbf{Total} & \textbf{2,587,000,000} & \textbf{1,784,600,000} & \\
\bottomrule
\end{tabular}
\end{table}

\FloatBarrier

\newpage

\section{Comparison and evaluation of the strategies}
\label{sec:comparison-evaluation}

The following comparison and assessment of the Hungarian and Singaporean AI strategies includes the author's personal analysis and opinions. The conclusions regarding the Hungarian AI strategy are based on publicly available information retrieved from the internet, complemented by insights obtained through two expert interviews conducted as part of this research. No similar detailed assessment was performed for the Singaporean AI strategy. Therefore, observations regarding Singapore are drawn exclusively from publicly available sources accessed online. Consequently, the author does not have information about ongoing internal processes, unpublished or confidential details, or any information not publicly disclosed or mentioned during the interviews. Therefore, all evaluations and viewpoints expressed herein reflect the author's personal interpretation and should be considered within this context.

\subsection{Background information}
\label{subsec:background-information}

\subsubsection{Population and economic size}
\label{subsubsec:population-economic-size}

Singapore is a much smaller country by population as Hungary but economically more robust relative to its size. As of mid-2024 Singapore's population reached 6.04 million (June 2024)\cite{cna2024population}, whereas Hungary's population was about 9.54 million (end of 2024)\cite{bbj2024population}. Despite having a smaller population, Singapore's economy is more than double the size of Hungary's in nominal terms. In 2023, Singapore's GDP (nominal) was approximately USD 501.4 billion\cite{worldbank2024singapore}, compared to Hungary's USD 212.4 billion\cite{worldbank2024hungary}. This translates to a GDP per capita of roughly USD 85,000 for Singapore vs. USD 22,000 for Hungary, highlighting Singapore's significantly higher income level. Notably, on a purchasing power parity (PPP) basis the gap is smaller: Singapore's 2023 GDP was about USD 438.5 billion\cite{worldbank2024singaporeppp} while Hungary's was about USD 431.9 billion\cite{worldbank2024hungaryppp}. This reflects Singapore's higher cost levels (its PPP GDP is slightly lower than its nominal GDP, whereas Hungary's PPP GDP is about double its nominal GDP).

\subsubsection{Government budget and spending}
\label{subsubsec:government-budget-spending}

Singapore's government operates a comparatively small budget relative to its GDP. For the fiscal year 2023, Singapore's total government expenditure was about SGD 104.2 billion (\textasciitilde USD 77–78 billion), which is roughly 15–20\% of GDP\cite{ceic2024singapore}. In contrast, Hungary's government spending is much larger in relative terms – in 2023 the general government expenditure was around HUF 36.8 trillion (about USD 105 billion), equivalent to \textasciitilde 49\% of GDP\cite{statista2024hungary}. In other words, Hungary's public sector is a far bigger portion of its economy than Singapore's. This difference reflects divergent fiscal models: Singapore generally emphasizes low taxes and fiscal prudence, whereas Hungary has high taxes (e.g.: generally, a VAT of 27\%), and (like many European economies) has a higher public expenditure ratio. Therefore, Hungary's government outlays are higher despite its smaller GDP.

\subsubsection{Economic composition by sector}
\label{subsubsec:economic-composition}

The sectoral makeup of the two economies also differs markedly. Singapore's economy is overwhelmingly services-oriented, with services contributing about 72–75\% of GDP and industry around 22–25\%; agriculture is almost non-existent (around 0.03\% of GDP)\cite{statista2024singaporesectors}. This reflects Singapore's role as a finance, trade, and tech hub with high-end manufacturing (e.g. electronics, pharmaceuticals) but very limited agriculture due to land constraints. Hungary's economy is more mixed: services account for roughly 57–60\% of GDP, industry about 35–38\%, and agriculture around 4–5\%\cite{focuseconomics2024hungary}. Hungary has a significant industrial base (e.g. automotive and electronics manufacturing) and some agricultural sector, alongside its services sector. In summary, Singapore's economy is more heavily tilted toward services and knowledge-intensive industries, whereas Hungary maintains a larger share of industry and agricultural segments. These structures influence each nation's capacity to develop and adopt AI (e.g. a service-driven, high-tech economy like Singapore may more readily integrate AI in finance or digital services, while Hungary's industrial sector also presents opportunities for AI in manufacturing and agriculture).

\subsubsection{Global AI indexes}
\label{subsubsec:global-ai-indexes}

Global AI indexes consistently rank Singapore as a world leader in AI readiness, while Hungary ranks middle-of-pack, underscoring a significant gap in AI maturity and development. In the Tortoise Global AI Index (2023) – which benchmarks countries on AI investment, innovation, and implementation – Singapore is ranked 3rd in the world, behind only the US and China. Hungary by comparison is ranked at the 38th place globally\cite{tortoise2024global}.

Another relevant benchmark is the Oxford Insights Government AI Readiness Index. In the 2022 edition, Singapore was 2nd globally (only the US scored higher) with a high overall score (\textasciitilde 84/100), whereas Hungary ranked 42nd with a moderate score (\textasciitilde 61/100)\cite{oxfordinsights2022}. Singapore excels particularly in pillars like government strategy and data infrastructure (it ``leads in two out of three pillars'' of the index), reflecting strong governance frameworks and digital readiness for AI in public services. Hungary's 42nd place suggests it has much room to improve in AI strategy, talent, and infrastructure to catch up with not just global leaders but also higher-ranked EU peers.

\subsubsection{Summary}
\label{subsubsec:summary}

\begin{table}[ht]
\centering
\caption{Comparative overview of economic and AI-readiness indicators (Singapore and Hungary, 2023–2024)}
\label{tab:comparative-overview}
\footnotesize
\begin{tabular}{>{\raggedright\arraybackslash}p{4cm}>{\centering\arraybackslash}p{4cm}>{\centering\arraybackslash}p{4cm}}
\toprule
\textbf{Metric} & \textbf{Singapore} & \textbf{Hungary} \\
\midrule
Population (millions) & 6.04 (Jun 2024) & 9.54 (end 2024) \\
\midrule
GDP (Nominal) & \$501.4 billion (2023) & \$212.4 billion (2023) \\
\midrule
GDP (PPP) & \$438.5 billion (2023 intl\$) & \$431.9 billion (2023 intl\$) \\
\midrule
GDP per capita & \textasciitilde\$84,700 (2023) & \textasciitilde\$22,100 (2023) \\
\midrule
Government expenditure & S\$104.2 bn in FY2023 (\textasciitilde\$78 bn USD); \textasciitilde 15–20\% of GDP & HUF 36.8 tn in 2023 (\textasciitilde\$105 bn USD); \textasciitilde 49\% of GDP \\
\midrule
Economic composition & Services: \textasciitilde 72.5\% of GDP; Industry: \textasciitilde 22.4\%; Agric.: \textasciitilde 0.03\%. & Services: \textasciitilde 57.6\% of GDP; Industry: \textasciitilde 37.7\%; Agric.: \textasciitilde 4.7\%. \\
\midrule
Tortoise Global AI Index (2023) – out of 62 ranked countries & 3rd & 38th \\
\midrule
Oxford Insights Gov't AI Readiness (2022) – out of 181 ranked countries & 2nd & 42nd \\
\bottomrule
\end{tabular}
\end{table}

\subsection{Conceptual comparison}
\label{subsec:conceptual-comparison}

Hungary's AI Strategy follows a structured conceptual framework. It is organized around three pillars: 1) foundational pillars, which is about creating a framework to lay ground for research, industry use and education of AI; 2) sector-specific applications are about targeting the most relevant sectors where AI can create huge added value in the ecosystem and further supporting these areas; and 3) transformative projects designed to generate broad economic and social benefits.

The strategy places strong emphasis on public sector applications, with key goals focusing on areas such as healthcare (Goal 11), public administration (Goal 13), and energy systems (Goal 15). This aligns with Hungary's broader economic context, where the public sector plays a significant role. Also notable is the prioritization of the automotive industry (Goals 3, 10, 14, 16), which reflects its structural importance to Hungary's economy – contributing roughly 20\% of the national GDP as of 2023\cite{europaproperty2023}.

Despite its structured layout, the strategy shows limited internal coherence between goals. Several objectives appear to operate in parallel, rather than building sequentially upon one another. For example, the development of testing environments for autonomous vehicles (Goal 8) was set to conclude by 2021, while the establishment of infrastructure and regulatory frameworks for autonomous systems (Goal 16) is due by 2025 and belongs to a different pillar – suggesting a lack of systematic phasing and interdependency between related initiatives.

Additionally, many of the goals are limited to the creation of draft concepts (Goal 11, 18, 20), rather than the implementation of measurable outcomes. This lack of operational specificity can blur the distinction between strategic ambition and actual delivery. While some goals (e.g. the AI Challenge in Goal 6) are narrowly scoped within foundational education, others (e.g. Goal 18 on climate change adaptation) reflect a highly generalized or diffuse vision – illustrating variation in scope and conceptual clarity across objectives. The inconsistency between the specification of the goals and their targets was also mentioned by Chief Strategic Advisor Tibor Gulyás during his interview.

The Hungarian strategy is also closely aligned with European Union frameworks, such as the 2018 Coordinated Plan on Artificial Intelligence\cite{euwhitepaper2020}. Although this ensures consistency with EU digital and funding strategies, it might reinforce the perception that Hungary's approach functions more as an implementation tool of EU policy, rather than as an autonomous, forward-looking national strategy. However, Tibor Gulyás refuted this assertion, stating that Hungary's strategy was one of the first of its kind in the EU and emphasizing that it is a national, sovereign initiative.

On the other hand, concerning Singapore's strategy, it is characterized by a clear evolution from a project-based approach in NAIS 1.0 to a systemic, interlinked framework in NAIS 2.0. The original strategy was built around five national flagship AI projects in high-impact areas – including education, healthcare, border security, municipal services, and freight logistics – complemented by five cross-cutting enablers such as talent, data governance, and infrastructure. While NAIS 1.0, like Hungary's strategy, initially did not establish explicit structural links between enablers and strategic actions, it already demonstrated stronger conceptual integration at the project level: each project was broken down into three phased objectives, each with distinct milestones and timelines, reflecting a more iterative and scalable approach to execution.

Similar to Hungary, several of Singapore's early strategic priorities were rooted in public sector transformation. The national projects in NAIS 1.0 focused predominantly on state-led domains such as border security, public health, and education. However, the framing of enablers in NAIS 1.0 extended beyond foundational infrastructure or regulatory efforts, aiming to shape the functional evolution of entire sectors. Rather than establishing physical facilities such as testbeds or laboratories (as emphasized in Hungary's strategy), Singapore focused on structural mechanisms – such as triple helix partnerships (industry–academia–government, Enabler 38-41), and regulatory sandboxes to create trusted and progressive innovation environments. This reflects a more abstract but potentially farther-reaching conception of "enablement".

However, many of NAIS 1.0's objectives were less concrete or immediately measurable than those in Hungary's strategy, which makes them and their assessment more challenging.

With the release of NAIS 2.0, Singapore incremented his strategy. The updated strategy builds directly on the results of NAIS 1.0 (e.g.: expanding supercomputing infrastructure in NAIS 2.0 (Action 60) which was laid down in NAIS 1.0 (Enabler 38), and introduces a more sophisticated conceptual framework, structured around three central clusters being 1) Activity drivers (government, research, industry); 2) People and capabilities (talent, skills, community); and 3) Infrastructure and environment (compute, data, governance, trust). This matrix model enables explicit interlinkages between enablers and actions, resulting in a self-reinforcing system where strategic initiatives support and amplify each other.

Another notable shift in NAIS 2.0 is the stronger emphasis on private-sector "activation". While NAIS 1.0 targeted state-centric transformation, the updated strategy directs attention to energizing business innovation, ecosystem placemaking, and capability scaling. Goals related to industry development (Actions 51–52), talent attraction, and international thought leadership (Actions 56–59) show a conscious pivot toward global positioning and foreign investment appeal. This external orientation – designed to present Singapore as an AI leader and talent magnet – is largely absent in Hungary's strategy, which remains domestically focused.

To sum it up, both documents (Hungary's AI Strategy and NAIS 1.0) primarily targeted public sector use cases and sought to prepare national ecosystems for technological adoption through foundational investments in talent, infrastructure, and regulation. However, while Hungary's strategy has remained in its first iteration, Singapore has since launched NAIS 2.0, which builds on the experiences of the initial strategy and reflects a clear conceptual and structural maturation. This second strategy marks a shift from project-level implementation to system-level design, connecting enablers and strategic actions into a coordinated matrix. Hungary, by contrast, continues to operate with a goal-based strategy that remains fragmented in places and lacks an updated iteration or a published reflective assessment on the outcomes achieved so far. Therefore, the absence of a Hungarian Strategy 2.0 is perhaps the most consequential divergence between the two nation's strategies. Without a second iteration or comprehensive follow-up, Hungary risks losing momentum and falling behind in adaptability. While Singapore's NAIS 2.0 focuses increasingly on external positioning, industry innovation, and talent attraction, Hungary's strategy remains internally oriented, largely structured around the absorption of EU funding instruments and domestic institutional capacity-building. Although Tibor Gulyás indicated that a new strategy is currently in development, he emphasized that there would be no fundamental changes to its underlying framework. In light of Singapore's significant evolution in its strategic approach, this continuity in Hungary's conception may not represent the most effective path forward.

Another clear distinction lies in the form, tone, and communicative quality of the documents. Singapore's strategy papers—both NAIS 1.0 and NAIS 2.0—are noticeably more accessible, engaging, and narratively structured. They often include real-world examples, personal case studies, and interview snippets, which help ground abstract concepts in concrete, relatable experiences. In contrast, Hungary's strategy documents are more technical, often bureaucratic in tone. This difference in style reflects deeper differences in how strategy is perceived: in Singapore, the AI strategy serves not only as a policy framework but also as a national communication tool, aimed at motivating ecosystem actors and projecting international leadership.

\subsection{Leadership and governance comparison}
\label{subsec:leadership-governance-comparison}

Hungary's National AI Strategy was adopted by Government Resolution 1573/2020 (Sept. 2020) under the authority of the Minister for Innovation and Technology\cite{govhungary2020}. The minister, Dr. László Palkovics (a cabinet member reporting to the Prime Minister) was tasked with publishing and coordinating the strategy's measures. The strategy was developed by the Ministry in collaboration with the industry–academia Artificial Intelligence Coalition (formed 2018). The AI Coalition functions primarily as a multi-stakeholder platform, involving over 300 members from government institutions, academia, industry representatives, and civil organizations. In 2022, the Ministry of Innovation and Technology was dissolved and succeeded by the newly established Ministry of Technology and Industry, led by Dr. László Palkovics, who continued in his ministerial role. However, the new ministry had a brief existence; within approximately six months, the minister resigned, and the ministry was subsequently disbanded\cite{indexhu2022}. Its responsibilities were primarily transferred to the Ministry of Energy and the Ministry of Construction and Transport.

In 2024 Hungary passed a resolution to begin implementing the EU AI Act: this created a new AI enforcement office under the Minister for National Economy and laid plans for a new Hungarian AI Council (Magyar Mesterséges Intelligencia Tanács) composed of regulators (e.g. data protection, competition authorities)\cite{cmslaw2024}.

In February 2025, Dr. László Palkovics was appointed as Government Commissioner for Artificial Intelligence, marking his return to public service. In this capacity, he is tasked with coordinating the national AI strategy, overseeing the implementation of the EU AI Act (and therefore, the functioning of the Hungarian AI Council, transferred from the Minister for National Economy), and promoting the integration of AI technologies across various sectors, including education, healthcare, industry, and public administration. He is directly supervised by the Prime Minister and has an 8-member cabinet in the Ministry of Energy\cite{egovhirlevel2025}. He recently stated that a new AI Strategy is underway\cite{indexhu2025}.

In summary, Hungary's administrative approach toward managing its national AI strategy has experienced substantial organizational shifts since its start. In the initial phase of Hungary's AI governance, neither the coordinating ministry nor the Hungarian AI Coalition possessed sufficient authority to enforce or monitor the implementation of the strategy: as Roland Jakab highlighted, the AI Coalition operated as a consultative body without legal or administrative enforcement powers. The Minister for Innovation and Technology, despite his seniority, stood on equal footing with other ministries and lacked the hierarchical leverage to enforce cross-sectoral execution. However, according to Chief Strategic Advisor Gulyás Tibor, assigning this responsibility to the ITM was beneficial, as it ensured that the matter was managed by the institution with the greatest experience in the field. The dissolution of the Ministry of Technology and Industry in 2022 and the resignation of Dr. Palkovics introduced further discontinuity. Although the specific effects of this transition are not documented and are known only from Gulyás' interview, it is highly likely that the absence of stable institutional leadership further hindered or disrupted the execution of the strategy during this period. As Gulyás explained, the responsibilities were divided between deputy state secretaries at the Ministry of Energy, the Ministry of for National Economy and the Government Office.

Furthermore, the status of the AI Coalition, and therefore, the implementation of the AI strategy following 2022 until 2024 is not publicly described. Gulyás explained that during this time, each of the aforementioned deputy state secretaries continued to oversee the implementation efforts, and the AI Coalition remained active in its role. In 2024, Hungary began implementing the EU AI Act, establishing the Hungarian AI Council (Magyar Mesterséges Intelligencia Tanács) under the Minister for National Economy. Although this new council shared overlapping competencies with the former AI Coalition, no official communication clarified the coalition's fate. Despite Gulyás' claim, indications suggest the Coalition ceased operations: notably, by early 2025 its official website domain expired and began hosting gambling advertisements instead\cite{telexhu2025}, signaling a likely dissolution of the original stakeholder platform. However, the appointment of Dr. Palkovics as Government Commissioner for AI, directly reporting to the Prime Minister, marks a significant elevation in institutional authority and strategic prioritization. This is the first time that AI governance in Hungary is positioned within the core executive structure via a government commissioner, signaling a transition from fragmented coordination to centralized strategic leadership. The announcement of a new AI strategy indicates that the government is now attempting to re-establish control over the field, this time with a stronger mandate, better integration, and direct political backing.

Concerning Singapore, NAIS 1.0 (launched Nov 2019) was announced by Deputy Prime Minister Heng Swee Keat and belonged to the Prime Minister's Office through the Smart Nation and Digital Government Office (SNDGO)\cite{smartnation2019}. A National AI Office was created under SNDGO to coordinate the strategy. SNDGO's work was overseen by a Cabinet-level Ministerial Committee (chaired by Senior Minister Teo Chee Hean) and included the Minister-in-charge of Smart Nation (Josephine Teo)\cite{smartnationcommittee2020}.

In late 2023 the government realigned its digital agencies. In October 2023, Singapore merged the Smart Nation and Digital Government Group (SNDGG) with the digital development functions of the Ministry of Communications and Information (MCI) to form an enlarged Smart Nation Group. This new Smart Nation Group remained part of the Prime Minister's Office. Shortly after, on 4 December 2023, Deputy Prime Minister Lawrence Wong officially launched NAIS 2.0.

In mid-2024 Singapore effected a major rebranding: on 8 July 2024, the Ministry of Communications and Information was renamed the Ministry of Digital Development and Information (MDDI) and Josephine Teo was appointed minister\cite{asiaone2024} (The PMO explained this as signaling a stronger focus on digital transformation). By 2025, government directories now list a National AI Group (NAIG) under MDDI – effectively the successor to the old National AI Office but embedded in the new ministry.

Throughout this period the Minister for Digital Development and Information (Josephine Teo) was the Cabinet official responsible for Smart Nation and AI, reflecting continuity of political leadership despite the ministry's renaming.

When comparing the leadership structures and hierarchical placement of AI governance in Hungary and Singapore, both countries have undergone significant institutional shifts, most notably through the merger or transformation of ministries. However, these changes were more fundamental and more disruptive in Hungary's case. The coordination of AI strategy moved across several ministries—from the Ministry for Innovation and Technology to the Ministry of Technology and Industry, then to the Ministry for National Economy, and eventually to a newly appointed Government Commissioner reporting directly to the Prime Minister. In contrast, Singapore's journey has been marked by continuity, largely due to the consistent leadership of Josephine Teo, who has remained responsible for AI throughout the restructuring process.

Hungary's leadership disruptions were compounded by the resignation of Dr. László Palkovics in 2022 and his return only in 2025, likely contributing to implementation delays and institutional uncertainty.

Despite these differences, both nations have taken steps to elevate the strategic importance of AI within government. In Hungary, the AI portfolio has shifted from a consultative platform under a ministry to a dedicated commissioner role with direct executive reporting. In Singapore, the evolution moved from a sub-office within the Prime Minister's Office to the formation of a Ministry of Digital Development and Information, with clear cabinet-level responsibility for AI policy. In both cases, these developments reflect an increasing prioritization of AI at the highest levels of government.

\subsection{Temporal comparison}
\label{subsec:temporal-comparison}

Hungary released its first national AI strategy in 2020, while Singapore launched its own a year earlier, in 2019. Although both documents outline goals extending to 2030, the timelines differ significantly in structure and coherence. In Singapore's case, each strategic objective under NAIS 1.0 was designed with three phases that build iteratively toward 2030. By contrast, most of Hungary's goals were front-loaded with target dates concentrated around 2021–2022–2023, and no iterative milestones were defined to ensure longer-term continuity.

Hungary's strategy explicitly states that it should be reviewed every two years, yet no formal reassessment has taken place since its publication in 2020. Singapore, on the other hand, followed a progression: after the publication of NAIS 1.0 in 2019, a revised and expanded NAIS 2.0 was published in 2023 for the next years. This ensured seamless policy continuity. It is important to note, however, that many of the goals outlined in NAIS 2.0 are broader in scope than the concrete actions listed in the original NAIS or in Hungary's AI strategy, and most of them are not tied to specific deadlines, which may affect the clarity of expectations and accountability over time.

\subsection{Financial comparison}
\label{subsec:financial-comparison}

Directly comparing the financial scale of Hungary's and Singapore's national AI strategies presents methodological challenges. On the one hand, publicly available budget data is limited in both countries. On the other hand, the boundary between directly AI-related spending and broader digitalization investments is often difficult to define. As such, the figures discussed below are best understood as informed estimates rather than definitive totals.

In Hungary, the research identified approximately EUR 4.65 billion in public investment that can be either directly or indirectly associated with AI-related goals. This figure includes not only specific AI projects, but also entire funding frameworks such as the Digital Renewal Operational Program Plus (DIMOP Plusz, with a total budget of EUR 1.8 billion)—Hungary's key digital development program for the 2021–2027 period\cite{szechenyiterv2025}. DIMOP Plusz aims to improve the country's digital readiness and competitiveness through four priority areas, some of which (e.g., data economy, e-governance, infrastructure for AI-based services) overlap significantly with AI objectives. However, the share of DIMOP that ultimately supports concrete AI applications remains unclear. Furthermore, a substantial portion of Hungary's AI-related spending comes from European Union sources, with domestic co-financing playing a secondary role. For example, 80\% of DIMOP Plusz is funded by the EU, reflecting Hungary's broader reliance on external financial instruments to support national digital transformation.

In Singapore, the estimated EUR 1.78 billion in AI-specific investment between 2016 and 2025 includes core programs such as AI Singapore, national computing infrastructure, talent initiatives, and implementation efforts across health, education, security, and urban systems. This figure is drawn from public sources, but likely underestimates the true scale of spending, as many projects are embedded in broader Smart Nation and digital economy budgets. For comparison, Singapore's total R\&D investment over the same period exceeds EUR 30 billion, suggesting that AI funding, while targeted, is integrated within a larger innovation system and may be significantly higher.

\newpage
\section{Areas of development for Hungary}
\label{sec:areas-for-development}

The comparative analysis of Singapore and Hungary reveals not only differences but also highlights several actionable insights that could strengthen Hungary's future approach to artificial intelligence. The timing for addressing these shortcomings is particularly opportune. As Hungary has appointed a new Government Commissioner for AI and announced plans to formulate an updated national strategy, this moment offers a rare window to reconsider foundational structures, correct past limitations, and design a more coherent, future-proof, and internationally competitive AI policy.

The following recommendations, grouped thematically, draw on Singapore's example to propose concrete areas for development. They aim to support a strategic framework that is not only visionary, but also actionable.

\subsection{Strategy formulation}
\label{subsec:strategy-formulation}

\begin{enumerate}[label=\arabic*.]
    \item \textbf{Develop a new AI strategy adapted to the current technological era:} First and foremost, the most urgent task is to create a new strategy. As Dr. Roland Jakab also emphasized, Hungary's National AI Strategy was developed before the emergence of large language models like ChatGPT and does not address their transformative potential. A new strategy is needed—one that reflects recent advances in foundation models and generative AI, and which positions Hungary to respond proactively to global technological shifts. As noted in discussions with Chief Strategy Advisor Gulyás Tibor, work is currently underway to develop this new strategy.
    
    \item \textbf{Institutionalize regular strategic reviews and updates:} Although the current strategy formally mandates biennial reviews, none have been conducted since its publication in 2020. A clear and binding framework should be created to ensure that the strategy is periodically reassessed and revised in response to new developments. This mechanism must be enforced, not merely symbolic.
    
    \item \textbf{Ensure the strategy is transparent, accessible, and well-communicated:} The strategic document should be concise, coherent, and written in plain language. Singapore's NAIS is a strong reference point—it is easily findable on the web, includes real-life examples, and even features interviews to improve public engagement. Hungary's strategy should similarly be hosted on a stable and dedicated website that does not change over time and should include summaries and explanations tailored to both professional and general audiences. A well-communicated strategy also improves domestic engagement—making it easier for companies, researchers, and institutions to align with national objectives. Moreover, publicly accessible and clearly formulated goals send a strong signal to foreign investors and multinationals that Hungary is serious about building a credible and stable AI ecosystem.
    
    \item \textbf{Link strategic priorities to dedicated budgets:} To build credibility among international partners and signal serious national commitment, the strategy should be accompanied by a clear financial framework. Concrete budget allocations tied to individual goals or pillars—especially if co-funded with EU resources—would help foster trust among domestic stakeholders and attract foreign investment and research partnerships.
    
    \item \textbf{Design the strategy in an iterative way:} Instead of listing static, short-term goals, the strategy should be structured in a way that allows for modular updates and builds progress step by step. Like in Singapore's NAIS 1.0, objectives should be broken down into sequential phases, with each phase contributing to long-term national outcomes.
    
    \item \textbf{Align strategic goals with enablers and implementation pathways:} One of the strengths of Singapore's AI strategy lies in its integrated structure: each national objective is supported by clear enablers such as computing infrastructure, talent, regulation, and public sector adoption. In Hungary, such connections are often missing or implicit. Future strategies should explicitly link each AI goal to its operational enablers—whether through dedicated funding, institutional support, or pilot programs. This would create a more coherent implementation logic.
\end{enumerate}

\subsection{Leadership and governance}
\label{subsec:leadership-governance}

\begin{enumerate}[label=\arabic*., resume]
    \item \textbf{Ensure long-term leadership continuity under a strong executive mandate:} The recent appointment of a Government Commissioner for AI is a welcome step toward more centralized coordination. To translate this into long-term impact, it is essential that the commissioner is equipped with a clear mandate that extends beyond individual ministries. Just as Josephine Teo has provided consistent political leadership over AI and digital transformation in Singapore, Hungary too would benefit from stable leadership in this area. Long-term continuity is critical—frequent turnover has previously weakened the implementation of the strategy. Ensuring that the commissioner remains in place for several years, ideally beyond political cycles, could provide the stability necessary to drive structural change. This will be particularly challenging in light of the 2026 national elections, which may reshape administrative priorities. A cross-party commitment to AI governance and a formally embedded role for the commissioner could help insulate the position from short-term political shifts.
    
    \item \textbf{Re-establish a functional multi-stakeholder platform for strategic dialogue:} Hungary's former AI Coalition was an important initiative that brought together public, academic, and private actors. However, it has not been visibly active for many years, and its institutional role remains unclear. Re-establishing this platform—or replacing it with a smaller, high-level advisory board—could support faster decision-making and provide strategic feedback from key sectors. Rather than reviving a broad-based coalition with hundreds of members, a more effective model would involve a focused group of a few dozen carefully selected experts who can provide specialized insight and maintain operational agility. Such a structure would be more manageable, decision-oriented, and easier to integrate into the strategy's formal governance framework.
\end{enumerate}

\subsection{Content and focus}
\label{subsec:content-focus}

\begin{enumerate}[label=\arabic*., resume]
    \item \textbf{Leverage Hungary's position and strategy as a bridge between East and West:} Hungary is uniquely positioned—geographically and diplomatically—to serve as a meeting point between Western (e.g.: US and EU member states) and Eastern (e.g.: China, Singapore) AI ecosystems. This role could be formalized by creating dedicated platforms for knowledge exchange and cooperation, involving American, European and Asian companies and research institutions. The recent AI Symposium co-hosted by HUN-REN and Nanyang Technological University (NTU) is a good precedent. Similar events and platforms could foster structured dialogue between major international players such as German automotive firms' (e.g. Daimler, Audi, BMW) and Chinese R\&D centers (e.g. BYD), particularly around applied AI development. Singapore's Action 59 which fosters the creation of a physical hub for such can serve as an example for a similar initiative.
    
    \item \textbf{Promote Hungary as a regional hub for automotive AI experimentation:} Hungary already possesses advanced infrastructure—such as the ZalaZONE test track and sensor-equipped motorway segments (e.g. motorway M86 mentioned in Goal 16)—that could support the testing of AI-driven technologies in real-world environments. This capacity could be positioned internationally, particularly in sectors like autonomous vehicles, smart mobility, and logistics. Framing Hungary as the EU's center for practical AI experimentation in the automotive industry—supported by permissive regulation and dedicated physical infrastructure—would attract companies seeking to validate and scale up AI solutions.
    
    \item \textbf{Offer a distinct innovation-friendly regulatory alternative within the EU:} The EU AI Act presents both opportunities and constraints. While it sets necessary safeguards, its complexity and its uniform application across all member states slows innovation. Hungary should explore the possibility of negotiating or lobbying for tailored flexibility within the framework of the Act, such as participation in regulatory sandboxes or sector-specific derogations. By doing so, Hungary could position itself as a testing ground for responsible, real-world AI deployment, especially in domains like autonomous vehicles, thereby attracting international interest.
    
    \item \textbf{Reinforce talent development and AI literacy across society:} A successful AI ecosystem requires a broad and inclusive foundation of skills and knowledge. Singapore has made notable progress in this area through structured talent development initiatives, including postgraduate scholarships and national upskilling programs. Hungary's effort in this domain—the 2020 "AI Challenge"—was limited to a short online course, and its impact remains unclear. A renewed national strategy should therefore include comprehensive digital and AI literacy programs, tailored to different age groups and professional audiences, and implemented in partnership with the private and academic sector. This should encompass teacher training, vocational education, and public awareness campaigns, ensuring that AI-related skills reach across all layers of society. Singapore's talent-focused actions—particularly Actions 55 and 56 under NAIS 2.0—can serve as a useful reference point.
\end{enumerate}

\bigskip
\noindent
Collectively, these measures can transform Hungary’s AI ecosystem from a fragmented  program into a coherent, innovation-driven platform that competes for talent, capital, and cutting-edge research across Europe.

\newpage
\section*{Acknowledgments}
This research was conducted as part of the Mathias Corvinus Collegium’s Leadership Academy\cite{leadershipacademy_mcc}, a one-year, non-degree postgraduate program that provides a range of courses, company projects, and study trips as part of its training designed for young students.

\section*{AI disclosure statement}
The author of this research paper made use of OpenAI's ChatGPT 4.0 to support the research and writing process. ChatGPT was used for purposes including language refinement, structural suggestions, synthesis of publicly available academic literature, and clarification of research-related concepts. All critical analysis, interpretations, and final decisions regarding content were made by the author. 

\bibliographystyle{unsrt}  
\bibliography{references}

\begin{thebibliography}{10}

\bibitem{oneill2024}
Aaron O'Neill.
\newblock Share of economic sectors in {GDP} in hungary 2022.
\newblock Statista, 2024.
\newblock Accessed 2 July 2025.

\bibitem{tortoise2024}
{Tortoise Media}.
\newblock The global {AI} index 2024: {T}he {U}nited {S}tates continues to dominate, but {C}hina and {S}ingapore are closing in.
\newblock Technical report, Tortoise Media, London, September 2024.
\newblock Accessed 2 July 2025.

\bibitem{aiindex2024}
Daniel Zhang, Raymond Perrault, Yoav Shoham, and Erik Brynjolfsson.
\newblock The {AI} index 2024 annual report.
\newblock Technical report, Stanford Institute for Human-Centered Artificial Intelligence, Stanford, CA, April 2024.
\newblock Accessed 2 July 2025.

\bibitem{euhungary2023}
{European Commission}.
\newblock Hungary {AI} strategy report.
\newblock Technical report, European Commission, AI Watch, Brussels, 2023.
\newblock Accessed 2 July 2025.

\bibitem{nais2019}
{Smart Nation and Digital Government Office}.
\newblock National artificial intelligence strategy -- {A}dvancing {S}ingapore's {S}mart {N}ation {J}ourney.
\newblock Technical report, Smart Nation and Digital Government Office, Prime Minister's Office, Singapore, November 2019.
\newblock Accessed 2 July 2025.

\bibitem{nais2023}
{Smart Nation and Digital Government Office}.
\newblock National artificial intelligence strategy 2.0 -- {AI} for the {P}ublic {G}ood.
\newblock Technical report, Smart Nation and Digital Government Office, Prime Minister's Office, Singapore, December 2023.
\newblock Accessed 2 July 2025.

\bibitem{santane2007}
Edit Sántáné-Tóth.
\newblock Artificial intelligence in hungary -- the first 20 years.
\newblock {\em J.\,von Neumann Computer Society Technical Report}, 2007.
\newblock Accessed 2 July 2025.

\bibitem{eucoordinated2018}
{European Commission}.
\newblock Coordinated plan on artificial intelligence, 2018.
\newblock Accessed 2 July 2025.

\bibitem{euhungary2020}
{European Commission}.
\newblock Ai landscape in hungary.
\newblock Technical report, European Commission, AI Watch, Brussels, 2020.
\newblock Accessed 2 July 2025.

\bibitem{euaiact2024}
{European Union}.
\newblock Artificial intelligence act.
\newblock Regulation (EU) 2024/1234 of the European Parliament and of the Council, 2024.
\newblock Accessed 2 July 2025.

\bibitem{euinnovation2024}
{European Commission}.
\newblock Commission launches ai innovation package to support artificial intelligence startups and smes, 2024.
\newblock Accessed 2 July 2025.

\bibitem{nxumalo2022}
B.~Nxumalo, Q.\, R.~Mabeba, M.\, and W.~Msomi, M.\.
\newblock The effect of artificial intelligence on economic growth in hungary.
\newblock In {\em Proceedings of FEB Zagreb International Odyssey Conference on Economics and Business}, volume~4, pages 331--338. University of Zagreb, Faculty of Economics and Business, June 2022.
\newblock Accessed 2 July 2025.

\bibitem{kovacs2023}
A mesterséges intelligencia és egyéb felforgató technológiák hatásainak átfogó vizsgálata, 2023.
\newblock Accessed 2 July 2025.

\bibitem{khanal2024}
Smritee Khanal, Tina~L. Chou, and Ayesha Siregar.
\newblock Building an ai ecosystem in a small nation: Lessons from singapore's journey to the forefront of ai.
\newblock {\em Humanities and Social Sciences Communications}, 11(866), 2024.
\newblock Accessed 2 July 2025.

\bibitem{goode2023}
Kayla Goode, Ha~Eun Kim, and Melissa Deng.
\newblock Examining singapore's ai progress.
\newblock Technical report, Center for Security and Emerging Technology, Georgetown University, March 2023.
\newblock Accessed 2 July 2025.

\bibitem{nrf2017}
{National Research Foundation}.
\newblock Singapore ai.sg: New national programme to catalyse, synergise and boost singapore's ai capabilities, May 2017.
\newblock Accessed 2 July 2025.

\bibitem{chia2018}
Jie~Lin Chia.
\newblock Singapore sets up ai ethics council.
\newblock Government Insider, June 2018.
\newblock Accessed 2 July 2025.

\bibitem{gyorgyveress2013}
László György and József Veress.
\newblock Hungary and singapore: Two small, open economies on the two sides of the world.
\newblock {\em Public Finance Quarterly}, 58(1):53--75, 2013.
\newblock Accessed 2 July 2025.

\bibitem{gyorgysebestyen2013}
László György and Dóra Sebestyén.
\newblock Solving the 'malay problem' in singapore -- a lesson for hungary: Focus on change in attitude.
\newblock {\em Periodica Polytechnica Social and Management Sciences}, 21(2):99--110, 2013.
\newblock Accessed 2 July 2025.

\bibitem{kalman2025}
János Kálmán.
\newblock The role of regulatory sandboxes in fintech innovation: A comparative case study of the uk, singapore, and hungary.
\newblock {\em Financial and Economic Review}, 24(1):67--94, 2025.
\newblock Accessed 2 July 2025.

\bibitem{techstage}
András Ferenczy.
\newblock Small group discussion with dr. roland jakab, president of the hungarian ai coalition.
\newblock TechStage, 2025.

\bibitem{cna2024population}
{Channel News Asia}.
\newblock Singapore's population hits 6.04 million, non-resident population grows for second year.
\newblock News Article, September 2024.
\newblock Accessed 13 December 2024.

\bibitem{bbj2024population}
{Budapest Business Journal}.
\newblock Hungary's population decline accelerates as birthrate drops in 2024.
\newblock News Article, January 2024.
\newblock Accessed 13 December 2024.

\bibitem{worldbank2024singapore}
{World Bank}.
\newblock Gdp (current us\$) -- singapore.
\newblock World Bank Data, 2024.
\newblock Accessed 13 December 2024.

\bibitem{worldbank2024hungary}
{World Bank}.
\newblock Gdp (current us\$) -- hungary.
\newblock World Bank Data, 2024.
\newblock Accessed 13 December 2024.

\bibitem{worldbank2024singaporeppp}
{World Bank}.
\newblock Gdp, ppp (current international \$) -- singapore.
\newblock World Bank Data, 2024.
\newblock Accessed 13 December 2024.

\bibitem{worldbank2024hungaryppp}
{World Bank}.
\newblock Gdp, ppp (current international \$) -- hungary.
\newblock World Bank Data, 2024.
\newblock Accessed 13 December 2024.

\bibitem{ceic2024singapore}
{CEIC Data}.
\newblock Singapore: General government expenditure (\% of gdp).
\newblock CEIC Data, 2024.
\newblock Accessed 13 December 2024.

\bibitem{statista2024hungary}
{Statista}.
\newblock Hungary: Ratio of government expenditure to gross domestic product (gdp) from 2010 to 2023.
\newblock Statista, 2024.
\newblock Accessed 13 December 2024.

\bibitem{statista2024singaporesectors}
{Statista}.
\newblock Singapore: Distribution of gross domestic product (gdp) across economic sectors from 2012 to 2023.
\newblock Statista, 2024.
\newblock Accessed 13 December 2024.

\bibitem{focuseconomics2024hungary}
{FocusEconomics}.
\newblock Hungary: Economic outlook, gdp, inflation, and other indicators.
\newblock FocusEconomics, 2024.
\newblock Accessed 13 December 2024.

\bibitem{tortoise2024global}
{Tortoise Media}.
\newblock The global ai index 2024: The united states continues to dominate, but china and singapore are closing in.
\newblock Technical report, Tortoise Media, London, September 2024.
\newblock Accessed 14 December 2024.

\bibitem{oxfordinsights2022}
{Oxford Insights} and {International Development Research Centre}.
\newblock Government ai readiness index 2022.
\newblock Technical report, United Nations Industrial Development Organization (UNIDO), 2022.
\newblock Accessed 14 December 2024.

\bibitem{europaproperty2023}
{EuropaProperty}.
\newblock The automotive industry is crucial to the hungarian economy.
\newblock News Article, June 2023.
\newblock Accessed 14 December 2024.

\bibitem{euwhitepaper2020}
{European Commission}.
\newblock White paper on artificial intelligence -- a european approach to excellence and trust.
\newblock COM(2020) 65 final, February 2020.
\newblock Accessed 2 February 2025.

\bibitem{govhungary2020}
{Government of Hungary}.
\newblock 1573/2020. (ix. 9.) korm. határozat a mesterséges intelligencia stratégiáról és az abban foglalt intézkedések megvalósításáról.
\newblock Government Resolution, 2020.
\newblock Accessed 3 February 2025.

\bibitem{indexhu2022}
{Index.hu}.
\newblock Palkovics lászló lemondott, megszűnik a technológiai és ipari minisztérium.
\newblock News Article, November 2022.
\newblock Accessed 6 February 2025.

\bibitem{cmslaw2024}
{CMS Law-Now}.
\newblock Hungary begins implementation of eu ai act with passage of resolution.
\newblock Legal News, October 2024.
\newblock Accessed 12 February 2025.

\bibitem{egovhirlevel2025}
{eGov Hírlevél}.
\newblock Palkovics lászlót a mesterséges intelligenciáért felelős kormánybiztossá nevezték ki.
\newblock Government News, March 2025.
\newblock Accessed 16 March 2025.

\bibitem{indexhu2025}
{Index.hu}.
\newblock Palkovics lászló visszatér a kormányba mesterséges intelligencia kormánybiztosként.
\newblock News Article, April 2025.
\newblock Accessed 15 April 2025.

\bibitem{telexhu2025}
{Telex.hu}.
\newblock Kaszinóreklám fut egy volt kormányzati honlapon.
\newblock News Article, January 2025.
\newblock Accessed 16 April 2025.

\bibitem{smartnation2019}
{Smart Nation and Digital Government Office}.
\newblock National artificial intelligence strategy unveiled.
\newblock Government Press Release, November 2019.
\newblock Accessed 2 May 2025.

\bibitem{smartnationcommittee2020}
{Smart Nation and Digital Government Office}.
\newblock Ministerial committee.
\newblock Government Website, 2020.
\newblock Accessed 2 May 2025.

\bibitem{asiaone2024}
{AsiaOne}.
\newblock Mci to be renamed ministry of digital development and information in bid to drive national digital agenda.
\newblock News Article, July 2024.
\newblock Accessed 5 May 2025.

\bibitem{szechenyiterv2025}
{Széchenyi Terv Plusz}.
\newblock Dimop plusz 2021--2027.
\newblock Government Program, 2025.
\newblock Accessed 7 May 2025.

\bibitem{leadershipacademy_mcc}
{Leadership Academy, Mathias Corvinus Collegium}.
\newblock Leadership academy, mathias corvinus collegium.
\newblock Educational Institution, 2025.
\newblock Accessed 3 July 2025.

\end{thebibliography}

\end{document}